\documentclass{article}
 \usepackage{amssymb,amsfonts,amsmath}

\DeclareMathOperator{\sgn}{sgn}
\DeclareMathOperator{\supp}{supp}

\newcommand{\<}{\langle}
\renewcommand{\>}{\rangle}
\newcommand{\R}{\mathcal{R}}
\newcommand{\Sc}{\mathcal{S}}

\newcommand{\F}{\mathcal{F}}

\newcommand{\B}{\mathcal{B}}
\newcommand{\C}{\mathcal{C}}
\newcommand{\Hc}{\mathcal{H}}

\newcommand{\J}{\mathcal{J}}

\newcommand{\M}{\mathcal{M}}
\newcommand{\Oc}{\mathcal{O}}
\newcommand{\T}{\mathfrak{T}}
\newcommand{\Di}{\mathcal{K}^{\scriptscriptstyle\Box}}
\newcommand{\Pg}{\mathcal{P}}
\newcommand{\Sb}{\Sc^{\scriptscriptstyle\Box}}
\newcommand{\sloc}{\mathrm{s.l.}}

\newcommand{\as}{\mathrm{as}}
\newcommand{\inc}{\mathrm{in}}
\newcommand{\out}{\mathrm{out}}
\newcommand{\ext}{\mathrm{ext}}
\newcommand{\al}{\alpha}
\newcommand{\be}{\beta}
\newcommand{\ka}{\kappa}
\newcommand{\ep}{\epsilon}
\newcommand{\ga}{\gamma}
\newcommand{\la}{\lambda}
\newcommand{\io}{\iota}

\newcommand{\mR}{\mathbb{R}}

\newcommand{\mC}{\mathbb{C}}
\newcommand{\1}{{\bf 1}}
\newcommand{\dsp}{\displaystyle}

\newcommand{\con}{\mathrm{const}}
\newcommand{\n}{\nabla}
\newcommand{\ti}{\widetilde}
\newcommand{\w}{\omega}
\newcommand{\W}{\Omega}

\newcommand{\ov}{\overline}

\newcommand{\wh}{\widehat}
\newcommand{\si}{\sigma}
\newcommand{\p}{\partial}
\newcommand{\dV}{\dot{V}}
\DeclareMathOperator{\Ker}{Ker} 
 
\DeclareMathOperator{\id}{id}

\title{Infrared problem and spatially local observables\\ in electrodynamics}
\author{Andrzej Herdegen%
\thanks{e-mail: herdegen@th.if.uj.edu.pl}\\
{\sl Institute of Physics, Jagiellonian University,}\\
{\sl Reymonta 4, 30-059 Cracow, Poland}}
\date{}
\begin{document}

\maketitle

\vspace{5ex}
\begin{abstract}
An algebra previously proposed as an asymptotic field structure in
electrodynamics is considered in respect of localization
properties of fields. Fields are `spatially local' -- localized in
regions resulting as unions of two intersecting (solid)
lightcones: a~future- and a~past-lightcone. This localization
remains in concord with the usual idealizations connected with the
scattering theory. Fields thus localized naturally include
infrared characteristics normally placed at spacelike infinity and
form a~structure respecting Gauss' law. When applied to the
description of the radiation of an external classical current the
model is free of `infrared catastrophe'.

%\vspace*{2ex} \noindent PACS numbers: 03.70.+k, 03.65.Bz, 11.10.-z
\end{abstract}
\vfill \eject

\sloppy
\renewcommand{\theequation}{\thesection.\arabic{equation}}

\section{Introduction}

The standard perturbative formulation of quantum electrodynamics,
effective as it is in predicting a~large scope of experimental
results, leaves many fundamental questions concerning the
structure of the theory unanswered. Among them is the one that
will concern us here, the so called ``infrared problem'' of
electromagnetic interaction (see Refs.\ \cite{jr,ms,haa,ste} for
discussion).

In short, the problem consists of difficulties in the theoretical
characterization of charged particles (and charged states) and the
construction of a~rigourous, and at the same time having
compelling quantum-mechanical interpretation, scattering theory in
electrodynamics. Its origin is the masslessness of the photon and,
consequently, the long-range character of the electromagnetic
interaction. In standard perturbational calculations it manifests
itself in the appearance of (IR-) divergencies, and at
intermediate stages is cured by the removal of the source of
trouble: either the photon is given a~small mass, or the
interaction is switched off in remote regions of spacetime.
However, the eventual removal of the regularizing parameter is
impossible unless previously an ``averaging over unobservable low
energy photons'' has been carried out. This may be an effective
calculational tool, but can hardly be regarded as a~real solution
of the problem.

Rigourous results on the issue are either incomplete or not widely
recognized as physically satisfactory, or both. The superselection
structure (related to the long-range behaviour) of prospective
quantum electrodynamics based on local observables  has been
investigated within the general algebraic approach \cite{bu82},
but a~consistent construction of a~model following the assumptions
has not been achieved so far, nor has a~charged particle been
given a~clear-cut, widely accepted characterization, despite some
tentative propositions in the direction \cite{bu91,haa}. At a~less
mathematically elevated level there have been attempts at
incorporating long-range aspects of electromagnetic interaction
into the dynamics of asymptotic fields \cite{kf,zw}. The results,
as assessed by Steinmann in his book on ``perturbative axiomatic''
electrodynamics, are ``too complicated to be of much practical use
either for the calculation of cross sections or for providing
insights into the underlying structures of the theory''
(\cite{ste}, p.\ 219), which is an opinion we share. However,
whether the approach proposed by Steinmann himself offers a~more
convincing alternative is open to debate: his analysis involves a
complicated transformation from the calculational
(indefinite-metric) representation to the physical one, and
eventually relies on a~somewhat arbitrary redefinition of the
cross-section for charged particles.

Summing up, the infrared problem is still open to further study,
which is a~task we take up here. We further develop the approach
to the infrared problem initiated in our earlier papers (Ref.\
\cite{her98} and papers cited therein; see also \cite{her05});
this proposition may be regarded as an attempt at an algebraic
formulation of an asymptotic dynamics respecting the long-range
nature of electromagnetic field and Gauss' law. The term
``asymptotic'' is meant here in the sense used in the scattering
theory, but the algebra is rather postulated then derived from
a~complete theory. In Ref.\,\cite{her98} the algebra was
formulated in terms of independent asymptotic variables (the
electromagnetic field part using the null infinity variables
similar to those used by other authors before, as discussed in
\cite{her98}; we do not use the conformal compactification
technique of Penrose). Here we extend our discussion by including
the issues of spacetime localization of fields and we formulate
anew the interpretational issues involved. At the center of our
discussion of localization stands an idea of spatial locality,
which we postulate and proceed to explain.

We first recall that one of the standard paradigms of quantum
field theory is spacetime locality of fundamental observables.
This is made most explicit in the algebraic approach based on a
net of algebras attached to bounded regions in spacetime
\cite{haa}; other physical observables may be approximated by
those strictly local. Strictly nonlocal effects, such as global
charges, are only to be found in the characteristics of the
\emph{states} on the algebra of local observables; locally the
global charge is irrelevant, one can always place a~compensating
charge ``behind the Moon''.

This picture is of course fruitful and seems to be well-founded in
the physical practice. However, physics is full of idealizations
contradicting, in strict sense, results or practice of experiment,
but theoretically helpful. One of such idealizations at the base
of the scattering theory (not only within quantum field theory) is
the notion of a~causally (``in'' or ``out'') asymptotic quantity
(asymptotic current, velocity, etc.). Strictly speaking, such a
quantity is beyond any physical experiment, which always takes a
finite time-span. Nevertheless, it is fruitful and helpful to
accept that if we wait long enough then the stabilization of the
result of a~measurement gives us information on a~quantity
extending unchanged (in appropriate sense) to infinite past or
future.

If we agree on this interpretation we can wonder why not include
in the fundamental structure of the theory observables with
localization extending to infinite past or future. We explore in
this paper consequences of this suggestion. We do not see the need
for including in the defining structure of the model localization
regions of infinite spatial extension -- the observables are
``spatially local''.

We give an outline of our ideas, ignoring most mathematical
subtleties, in Section \ref{ideas}. Section \ref{test} contains
the construction of various test functions spaces needed for a
rigourous construction of our algebraic model of asymptotic
fields, which is discussed in Section \ref{alg}. Section
\ref{scat} explains the expected role of the model in the
scattering theory and illustrates the idea on a~simple example of
radiation by a~classical current. Appendices contain some
technical material. The Lorentz product is denoted by $x\cdot y$
and has signature $(+,-,-,-)$. The spacetime integration element
is denoted $dx$ etc.

\pagebreak

\setcounter{equation}{0}

\section{Main ideas}\label{ideas}

To place our analysis in context we review well-known structures
appearing in the standard quantization of free electromagnetic
fields. This quantization is best described with the use of a
symplectic structure on the space of test fields. Let first our
space be the space of free fields satisfying Maxwell equations in
the whole spacetime, having compact support on each Cauchy
surface. The symplectic form is then supplied by
\begin{equation}\label{sfc}
 \{ F_1, F_2\}_C= \frac{1}{4\pi}
 \int_{\Sigma} (F_1{}^{ab}A_2{}_b-F_2{}^{ab}A_1{}_b)(x)\, d\si_a(x)\,,
\end{equation}
where integration extends over a~Cauchy surface $\Sigma$ and
$d\sigma_a$ is the dual integration form on $\Sigma$. This
symplectic  form is hypersurface- and gauge-independent.
In~particular, if a~spacelike hyperplane $t=\con$ in a~given
Minkowski frame is chosen, and the potentials are in radiation
gauge, then the form becomes
\begin{equation}\label{sfr}
 \{\vec{E}_1,\vec{A}_1; \vec{E}_2,\vec{A}_2\}_R=
 \frac{1}{4\pi}\int
 [\vec{E}_1\cdot\vec{A}_2-\vec{E}_2\cdot\vec{A}_1](t,\vec{x})\,d^3x\,.
\end{equation}
In this representation the electromagnetic field is represented by
 a~pair of 3-di\-ver\-gence-free fields $\vec{E}$, $\vec{A}$
supplying its initial data, which has the advantage of uniqueness.
However, this representation is inconvenient for the discussion of
spacetime localization, so another transform of (\ref{sfc}) is
needed.

If the potentials satisfy Lorenz condition, then one easily shows
by using Stokes' theorem that (\ref{sfc}) may be written as
\begin{equation}\label{sfaa}
  \frac{1}{4\pi}
  \int_{\Sigma} (\n^aA_1^b\,A_2{}_b-\n^aA_2^b\,A_1{}_b)(x)\,
  d\si_a(x)\,.
\end{equation}
Let now the potential $A_1$ be represented as the radiation (i.e.,
as usual, the retarded minus the advanced) potential of a
conserved, smooth, compactly supported current $J_1$, that is
\begin{equation}\label{rad}
    A_1^a(x)=4\pi\int D(x-y)J_1^a(y)\,dy\,,
\end{equation}
where $D(x)=\frac{1}{2\pi}\sgn(x^0)\delta(x^2)=-D(-x)$ is the
Pauli-Jordan function. Using this in (\ref{sfaa}) and noting that
$A_2$, being a~solution of the wave equation, may be expressed in
terms of initial data as
\begin{equation}\label{ic}
    A_2^b(y)=\int_\Sigma [D(y-x)\n^aA_2^b(x)+\n^aD(y-x)A_2^b(x)]\,
    d\sigma_a(x)\,,
\end{equation}
we find that (\ref{sfc}) may be written as
\begin{equation}\label{ja}
    \int J_1\cdot A_2(y)\,dy\,.
\end{equation}
This again is gauge-independent, so the Lorenz condition on $A_2$
may be dropped. Thus we find, by antisymmetry, that for smooth,
compactly supported currents the symplectic structure (\ref{sfc})
has another representation
\begin{equation}\label{sfj}
    \{J_1,J_2\}=\frac{1}{2}
    \int [J_1\cdot A_2-J_2\cdot A_1](y)\,dy\,,
\end{equation}
where $A_i$ are any potentials of the fields $F_i$ produced as
radiation fields by the currents $J_i$. If $A_i$ are chosen as in
(\ref{rad}) then for local currents $J_i$ this becomes
\begin{equation}\label{sfcc}
    4\pi\int J_1^b(x)D(x-y)J_2{}_b(y)\,dx\,dy\,.
\end{equation}
The symplectic space is now formed by equivalence classes of
smooth, compactly supported conserved currents producing the same
electromagnetic radiation field.

Finally, one notes, that for each such current $J^a$ there is a
compactly supported 2-vector $\varphi^{ab}$ such that
$J^a=2\n_b\varphi^{ab}$ (which is the consequence of the
Poincar\'e lemma for the 3-form dual to $J^a$). With the use of
Gauss' theorem expression (\ref{ja}) is rewritten as
\begin{equation}\label{fif}
    \int\varphi_1^{ab}F_2{}_{ab}(x)\,dx
\end{equation}
and the symplectic structure acquires still another form
\begin{equation}\label{sffi}
    \{\varphi_1,\varphi_2\}_\varphi=\frac{1}{2}
    \int[\varphi_1^{ab}F_2{}_{ab}
    -\varphi_2^{ab}F_1{}_{ab}](x)\,dx\,.
\end{equation}
The symplectic space is now formed by equivalence classes of
smooth, compactly supported 2-vectors $\varphi$ producing the same
electromagnetic radiation field.

The quantization of the electromagnetic field consists,
heuristically, in the replacement of one of the variables in the
symplectic form by a~``quantum variable'' and imposition of the
commutation rule:
\begin{equation}\label{com}
    \big[\{F_1,F\}_C,\{F_2,F\}_C\big]=i\{F_1,F_2\}_C\,,
\end{equation}
where $F$ symbolizes the quantum field and $F_i$ are test fields.
Any of the discussed above forms of the symplectic structure may
be used instead. For the spacetime localization of variables forms
(\ref{sfj}) or (\ref{sffi}) are suited. We choose the former, as
the test fields have very clear physical interpretation in that
case: they are conserved electromagnetic currents used to probe
the quantum field. The tested field will be denoted $A(J_1)$, and
according to (\ref{sfj}) the intuitive content of this symbol is
\begin{equation}\label{smear}
    A(J_1)=\frac{1}{2}\int [J_1^a(x)A_a(x)-J^a(x)A_1{}_a(x)]\,dx\,,
\end{equation}
with $J_a(x)$ related to $A_a(x)$ as in (\ref{rad}). For local
$J_1$ this may be put in the form given by (\ref{ja}):
\begin{equation}\label{smearloc}
    A(J_1)=\int J_1^a(x)A_a(x)\,dx\,,\qquad\quad J_1\ \
    \mathrm{local}\,.
\end{equation}
If this intuition is to be confirmed, the spacetime localization
of $A(J_1)$ should be confined to the support of $J_1$.  The
quantization condition becomes
\begin{equation}\label{comj}
    [A(J_1),A(J_2)]=i\{J_1,J_2\}\,,
\end{equation}
which is consistent with the assumed localization and relativistic
causality. All this is, of course, standard, possibly except for
the use of currents $J_1$ instead of 2-vectors $\varphi_1$ as test
fields, and consequently $A(J_1)$ instead of
$F(\varphi_1)=\int\varphi_1^{ab}(x)F_{ab}(x)\,dx$ as smeared
fields.

We want to remind the reader another point concerning the
commutation relations (\ref{comj}). The symplectic product on the
rhs of this equation may be interpreted as $A_2(J_1)$. If one
introduces Weyl operators
\begin{equation}\label{wa}
    W(J_1)=\exp[-iA(J_1)]\,,
\end{equation}
then (\ref{comj}) may be written as
\begin{equation}\label{comwa}
    A(J_1)W(J_2)=W(J_2)[A(J_1)+A_2(J_1)]\,.
\end{equation}
This has a~clear interpretation: $W(J_2)$, when acting on a~vector
state, produces the radiation field due to the current $J_2$. It
is important to note, that the coefficient in the exponent
defining $W(J_2)$ is fixed by this relation. The commutation
relations may be rewritten as a~Weyl algebra:
\begin{equation}\label{weyl}
    W(J_1)W(J_2)=\exp[-\tfrac{i}{2}\{J_1,J_2\}]W(J_1+J_2)\,.
\end{equation}

The vacuum Fock representation of the commutation relations
(\ref{comj}) is usually discussed in textbooks as \emph{the}
theory of free electromagnetic fields. However, if so, this theory
is a~rather poor one: it is not wide enough to include
infrared-singular fields -- with the long-range tail of the type
produced in scattering of charged particles. Several kinds of
response to this criticism are usually offered. One can argue that
all physics is done locally, so even neglecting remote
contributions one can approximate every physical situation. This
is the point of view adopted in the standard perturbational
electrodynamics; it goes together with the pragmatical approach to
the charged particle, as mentioned above, and is fundamentally not
convincing. If one accepts the real existence of the difficulty
one can still retain the scope of algebra (\ref{comj}), but
explore all ``physically reasonable'' representations; the problem
then is an overabundance of those: every distribution of
electromagnetic flux at spatial infinity labels a~different
representation. We think that it is legitimate to wonder whether,
indeed, all those labels are superselected with respect to
\emph{all} physically admissible observables. Here is the place,
logically, of attempts to introduce some ``variables at (spatial)
infinity'' into electrodynamics, as those of Refs.\ \cite{gzw80}
or \cite{sta89} (see also \cite{her05} for an account of
Staruszkiewicz's model). These attempts go in a~sense against the
paradigm of locality. It may be argued that all quantities should
be obtainable as limits of local ones, so one should not introduce
``by hand'' variables escaping such limiting process. We only
partly subscribe to this view, inasmuch as arbitrariness is
concerned. However, below we want to argue that the interpretation
of (\ref{comj}) described above naturally leads to the extension
of the algebra of commutation relations. But first, as we want to
treat some aspects of the interaction of the electromagnetic field
with charged matter -- for definiteness: electrons and positrons,
to fix notation we briefly summarize the quantization of the Dirac
field.

For the space of free Dirac fields with compact support on Cauchy
surfaces there is an invariant scalar product
\begin{equation}\label{dc}
    \<\psi_1,\psi_2\>_C=\int_\Sigma\ov{\psi_1}\gamma^a\psi_2(x)\,d\sigma_a(x)\,,
\end{equation}
which makes the space into a~pre-Hilbert space. Let the field
$\psi_1$ be obtained as
\begin{equation}\label{psch}
    \psi_1(x)=\frac{1}{i}\int S(x-y)\chi_1(y)\,dy\,,
\end{equation}
where $\chi_1$ is a~smooth, compactly supported 4-spinor field,
and $S(x)$ is the standard Green function of the free Dirac field:
\begin{equation}
 S(x)=(i\ga\cdot\p+m)D(m,x)\,,\quad
 D(m,x)=\frac{i}{(2\pi)^3}
 \int\sgn p^0\delta(p^2-m^2)e^{-ip\cdot x}dp\,.
\end{equation}
 Putting (\ref{psch}) into (\ref{dc}) and using the initial data
problem solution for $\psi_2$:
\begin{equation}\label{idd}
    \psi_2(y)=\frac{1}{i}\int_\Sigma S(y-x)\gamma^a\psi_2(x)\,d\sigma_a(x)
\end{equation}
one finds that (\ref{dc}) may be written as
\begin{equation}\label{chps}
    \int \ov{\chi_1(y)}\psi_2(y)\,dy\,.
\end{equation}
If all free Dirac fields are represented as in (\ref{psch}) then
our test fields space consists of equivalence classes of smooth
compactly supported $4$-spinor fields producing the same free
Dirac field, with the pre-Hilbert structure given by the product
\begin{equation}\label{preh}
    \<\chi_1,\chi_2\>=
    \frac{1}{i}\int\ov{\chi_1(x)}S(x-y)\chi_2(y)\,dx\,dy\,.
\end{equation}
The quantized Dirac field is now a~quantum variable $\psi(\chi_1)$
depending anti-linearly on $\chi_1$ and satisfying relations
\begin{equation}\label{antcom}
    [\psi(\chi_1),\psi(\chi_2)]_+=0\,,\qquad
    [\psi(\chi_1),\psi(\chi_2)^*]_+=\<\chi_1,\chi_2\>\,.
\end{equation}

In standard perturbational electrodynamics of interacting fields
one starts with the approximation of completely decoupled
electromagnetic and Dirac fields. However, one has to remember,
that electrodynamics is a~constrained theory, in which Gauss' law
should hold. Therefore the electromagnetic field of this
approximation is qualitatively different from the total field, and
the contradiction between locality and Gauss' constraint pervades
much of more conscious approaches to the problem, with some
complicated and physically not quite uncontroversial
transformations ``from local to charged sectors'' appearing at
late stages. We are of the opinion, that one should remove
contradiction at the starting point, even at the expense of
locality. Our aim is thus the replacement of the uncoupled algebra
of the electromagnetic and Dirac fields by some modified
asymptotic algebra, taking correctly into account the infrared
aspects of the electromagnetic field and Gauss' law. In the
asymptotic, ``in'' or ``out'' region, the interaction is weak, but
the long-range structure should survive.

We return first to the electromagnetic part of the algebra. Recall
that $A(J_1)$ is the field tested by a~conserved current $J_1$. It
is true that an idealization assuming currents extending to
spatial infinity does not seem to be justified. However, other
currents of noncompact support are commonplace in physics. More
than that, no charged current can be truncated to vanish at late
or early times, and this offers an example of a~physical quantity,
which cannot be approximated by quantities supported locally in
spacetime. Currents normally carried by scattered matter may have
compact support in spacelike directions, but for timelike
directions the typical asymptotic falloff is
\begin{equation}\label{falloff}
   J(\lambda x)\sim \lambda^{-3}J_\as(x)\,,\quad x^2>0\,,
   \quad \lambda\to\infty\,,
\end{equation}
which defines $J_\as$ as a~homogeneous of degree $-3$ asymptote of
$J$, supported inside the future and past lightcones. Moreover, in
all physical scattering situations there is
\begin{equation}\label{elas}
    x\wedge J_\as(x)=0\,,
\end{equation}
which reflects the fact that $J_\as$ is only due to asymptotically
free matter carrying electric charge (no magnetic charges).

Accordingly, we shall admit as the space of test fields a~class of
conserved currents of the type carried by charged matter, moving
freely at early and late times, but having compact support in
spacelike directions. As we shall see, this is sufficient to
result in the appearance of infrared degrees of freedom in the
algebra. This should not come as a~surprise if one recalls
classical analogues: an infrared singular field, however
low-energetic, induces a~finite phase change of the wave function
of a~quantum particle, and also causes an adiabatic shift of the
trajectory of a~classical charged particle \cite{sta81,her95}. In
both cases the size of the effect depends on the infrared
characteristic of the electromagnetic field, which shows that
physically typical currents are able to ``test'' those aspects of
the field.

We supplement, however, this extension of the test space with the
following restriction. The test currents space will slightly
differ for the ``in'' and ``out'' cases. In the former only the
past asymptotics, and in the latter the future asymptotics
admitted by (\ref{falloff}) may be different from zero. This seems
rational from the point of view of the regions in which those
fields are ``tested''. On the other hand, as we shall see, each of
these classes of currents produces the same class of radiation
fields (and the same as that obtained without these restrictions
on asymptotics), so in each case the role of $W(J_1)$ is the same.
We note that the restrictions automatically imply that the test
currents are globally charge-free (but may carry nonzero charges
in different asymptotic directions).

With the test space of currents thus extended, we shall have to be
more cautious with the symplectic form (\ref{sfj}). As we shall
see, this form is not completely gauge-independent any more on the
enlarged space. Therefore from now on we put for $A_i$ in this
form radiation potentials obtained according to (\ref{rad}). The
integrand in (\ref{sfj}) will be thus specified, and absolutely
integrable. On the other hand, the integrand in the double
integral of (\ref{sfcc}) will not be absolutely integrable in
general, so this form will not be used. After this specification
the symplectic form will become unambiguous for charge-free test
currents.

We now add charged particles.  We assume that the fields interact
only weakly, but we want to construct for this situation a~closed
algebra. The only remnant of the interaction which we take into
account is the fact that free charged particles carry their
Coulomb fields. Thus the quantum variables will now be interpreted
as:
\begin{equation}\label{inter}
 \begin{split}
    &\psi(\chi_1)\qquad \text{--- free charged field carrying its
    Coulomb field,}\\
    &A(J_1)\qquad \text{--- total electromagnetic field.}
 \end{split}
\end{equation}
For $A$~and $\psi$ separately we retain previous commutation
relations, but the above interpretation implies that these
variables should not be assumed to commute with each other, one
should expect a~relation of the intuitive form
\mbox{$A\,\psi=\psi\,[A+$ Coulomb field carried by $\psi]$}.
Recall once more that $A(J_1)$ is loosely $\{J_1,J\}$. Moreover,
with the use of Fourier-transformed fields
\begin{equation}\label{ft}
    \wh{\chi_1}(p)=\frac{1}{(2\pi)^2}\int\chi_1(x)e^{ip\cdot x}dx
\end{equation}
we have $\psi(\chi_1)=\int \ov{\wh{\chi_1}}(p)\wh{\psi}(p)\,dp$,
with $\wh{\psi}(p)$ describing a~particle with charge $-e$ moving
with the momentum $p$. Therefore we postulate the relation
\begin{equation}\label{apsi}
    A(J_1)\wh{\psi}(p)=\wh{\psi}(p)[A(J_1)-\{J_1,J_{p/m}\}]\,,
\end{equation}
where $J_v$ is the current connected with the particle with charge
$e$ moving freely with four-velocity $v$; remember that
$\wh{\psi}$ is supported on the mass hyperboloid. This current is
non-radiating, but on the other hand it differs from test currents
of the ``in'' and ``out'' space by having non-vanishing both
asymptotes.

\emph{A priori}, one has a~potential difficulty in the relation
(\ref{apsi}): a~particle with fixed momentum is completely
delocalized, so there is an ambiguity in the current~$J_{p/m}$.
It~turns out, however, that taking for $J_v$ the current of a
point particle moving along any straight line parallel to $v$ one
obtains the same value of $\{J_1,J_v\}$, depending only on the
long-range tail of the potential produced by $J_1$. We rewrite
(\ref{apsi}) in the form
\begin{equation}\label{wepsi}
    W(J_1)\psi(\chi_1)=\psi(S_{J_1}\chi_1)W(J_1)\,,
\end{equation}
where
\begin{equation}\label{s}
    \wh{S_{J_1}\chi_1}(p)=e^{-i\{J_1,J_{p/m}\}}\,\wh{\chi_1}(p)
    \quad\text{for}\quad p^2=m^2 \,.
\end{equation}

There are two important points to be made. Recall that the test
spinors for the Dirac field were assumed smooth and compactly
supported. This turns out to be inconsistent with the above
relation -- if $\chi_1$ is compactly supported, then
$S_{J_1}\chi_1$ is not. Therefore, similarly as in the
electromagnetic case, we have to extend the test function space.
We shall find that it is possible to choose these fields as
compactly supported in spacelike directions and decaying
polynomially in the timelike directions; the degree of the decay
may be chosen arbitrarily high without changing the element
$\psi(\chi_1)$.

The second point concerns gauge invariance. Consider element
$W(J_1)$ with $J_1$ producing pure gauge potential. As stated
above, the symplectic form is unambiguously defined for the
currents in one of the test classes (``in'' or ``out''), so this
element commutes with electromagnetic field. However, it does not
commute with the Dirac field and (\ref{wepsi}) gives in that case
\begin{equation}\label{gau}
    W(J_1)\psi(\chi_1)=e^{i\Lambda_1e}\psi(\chi_1)W(J_1)\,,\qquad
    A_1\ \text{pure gauge}\,,
\end{equation}
where the scalar $\Lambda_1$ is determined by the infrared
characteristic of the Lorentz potential $A_1$ produced by $J_1$.
Thus for such $J_1$ the element $W(J_1)$ should be interpreted as
$\exp[-i\Lambda_1Q]$, with $Q$ -- the total charge observable. The
quantization of charge in units of $e$ means that $\Lambda_1e$
should be interpreted as a~phase variable, which we shall take
into account below.

The elements $W(J_1)$ and $\psi(\chi_1)$ satisfying relations
(\ref{weyl}), (\ref{antcom}) and (\ref{wepsi}) form our algebra.
In the next section we give precise meaning to test fields of
these generating elements. Section \ref{alg} gives then a~precise
formulation of the algebra.

\setcounter{equation}{0}

\section{Test functions spaces}\label{test}

The geometry of the spacetime is given by the affine Minkowski
space $\M$. If a~ref\-erence point $O$ is chosen, then each point
$P$ in $\M$ is represented by a~vector $x$ in the associated
Minkowski vector space $M$ according to $P=O+x$. We mostly keep
$O$ fixed and use this representation, but we also remember to
control the independence of structures from $O$. If a~Minkowski
basis $(e_0,\ldots,e_3)$ in $M$ is chosen, then we denote
\mbox{$x=x^ie_i$}. We also then use the standard multi-index
notation $x^\alpha=(x^0)^{\alpha_0}\!\ldots (x^3)^{\alpha_3}$,
\mbox{$|\alpha|=\alpha_0+\ldots+\alpha_3$},
$D^\beta=\p_0^{\beta_0}\!\ldots\,\p_3^{\beta_3}$, where
$\p_i=\p/\p x^i$. We associate with the chosen Minkowski basis a
Euclidean metric with unit matrix in that basis, and denote $|x|$
the norm of $x$ in that metric. For a~tensor or spinor
$\C^\infty(\M)$ field $\phi$ we introduce for each $\ka\geq0$ and
$l=0,1,\ldots$ a~seminorm
\begin{equation}\label{semi}
    \|\phi\|_{\ka,\,l}=\sup(1+|x|)^\ka|D^\beta \phi_j(x)|\,,
\end{equation}
where supremum is taken over $x\in M$, all $\beta$ such that
$|\beta|=l$, and $j$ running over component indexes of the field
in the chosen basis. For fixed $l$ the seminorms form an
increasing net over $\ka\geq0$. For fixed $\ka$ and $l$ seminorms
$\|.\|_{\ka,l}$ and $\|.\|'_{\ka,l}$ associated with two reference
systems $(O,(e_i))$ and $(O',(e'_i))$ are equivalent. For fixed
$O$ this follows from the equivalence of norms $|x|$ and $|x|'$
and from the linearity of components transformations, while for
the translation $O'=O+a$ from the estimate
$(1+|x-a|)^\ka\leq\con(\ka,a)(1+|x|)^\ka$. If one denotes
$\phi_a(x)=\phi(x-a)$ then it follows
\begin{equation}\label{transl}
    \|\phi_a\|_{\ka,\,l}\leq\con\,\|\phi\|_{\ka,\,l}\,.
\end{equation}

Seminorms (\ref{semi}) are used in this section to construct the
spaces $\J_\as$ and $\Di$ which will supply test functions for
elements $W(J)$ and $\psi(\chi)$ respectively. We also equip the
space $\J_\as$ with the natural topology, although it will not be
used in this paper. Space $\Di$, being a~subspace of the Hilbert
space of the scalar product (\ref{preh}), inherits its topology.
The reference for inductive limit spaces are the books \cite{koe}
and \cite{rob}.

\subsection{Spaces $\boldsymbol{\Sc_{\ka+}}$}

Consider the space $\C^\infty(\M)$ of fields of a~given geometric
type -- not to burden notation this type will be kept implicit.
For each $\ka>0$ we define the subspace
\begin{equation}\label{kappa}
    \Sc_\ka=\big\{\phi\in\C^\infty\mid
    \|\phi\|_{\ka+l,\,l}<\infty,\, l=0,1,\ldots\big\}\,.
\end{equation}
With the topology determined by the family of seminorms which
define them, these spaces are locally convex, Fr\'echet spaces,
independent of the choice of a~reference system $(O,(e_i))$. We
denote the topology of $\Sc_\ka$ by $\T_\ka$. For each $\ka$ the
net of spaces $\Sc_{\ka+\ep}$, $\ep\in(0,1)$, is decreasing, so
the union
\begin{equation}\label{inli}
    \Sc_{\ka+}=\bigcup_{0<\ep<1}\Sc_{\ka+\ep}
\end{equation}
is a~vector space, a~subspace of $\Sc_\ka$. For $\ep>\ep'$ the
natural embedding $\Sc_{\ka+\ep}\mapsto\Sc_{\ka+\ep'}$ is
continuous. Denote by $\T_{\ka+}$ the strongest locally convex
topology on $\Sc_{\ka+}$ in which all natural embeddings
$\Sc_{\ka+\ep}\mapsto\Sc_{\ka+}$ are continuous. It is easy to see
that this topology is stronger than the topology induced on
$\Sc_{\ka+}$ by the topology $\T_\ka$. Therefore $\T_{\ka+}$ is
Hausdorff and we can conclude that the space
$(\Sc_{\ka+},\T_{\ka+})$ is the topological inductive limit of the
spaces $(\Sc_{\ka+\ep},\T_{\ka+\ep})$. The relations between the
spaces and topologies are summarized by
\begin{equation}\label{spatop}
    \Sc_{\ka+\ep}\subset
    \Sc_{\ka+}\subset\Sc_\ka\,,    \quad
    \T_{\ka+\ep}>\T_{\ka+}>\T_\ka\,,
\end{equation}
where the sign $>$ means that the topology to the left is stronger
than the topology induced by the one to the right.

\subsection{Spaces $\boldsymbol{\Sc^\ka_{\ka+}}$}

Consider the homogeneity operator $H=x\cdot\p$ and denote
$H_\ka=H+\ka\id$ for $\ka>0$. If $H_\ka f(x)=0$ then $f$ is
homogeneous of degree $-\ka$, so $H_\ka$ is injective
on~$\C^\infty$. One has $\dsp|D^\beta
H_\ka\phi(x)|\leq\sum_{i=0}^3|x^i||\p_iD^\beta\phi(x)|
+(\ka+|\beta|)|D^\beta\phi(x)|$, so~for $\phi\in\Sc_{\ka+\ep}$
there is
\begin{equation}\label{hphi}
    \|H_\ka\phi\|_{\ka+\ep+l,\,l}
    \leq\con\,\Big(\|\phi\|_{\ka+\ep+l+1,\,l+1}
    +\|\phi\|_{\ka+\ep+l,\,l}\Big)\,.
\end{equation}
Conversely, let now $\psi\in\Sc_{\ka+\ep}$ and define
\begin{equation}\label{hinv}
    \phi(x)=\int_0^1u^{\ka-1}\psi(ux)du\,.
\end{equation}
Using the bounds on $D^\beta\psi(x)$ one finds that
differentiation may be pulled under the integral sign
\begin{equation}\label{hinvd}
    D^\beta\phi(x)=\int_0^1u^{\ka+|\beta|-1}[D^\beta\psi](ux)\,du\,,
\end{equation}
and then it is easily seen that $H_\ka\phi=\psi$. The estimation
of the integral gives
\begin{equation}\label{phiest}
    |D^\beta\phi(x)|\leq\con\|\psi\|_{\ka+\ep+|\be|,|\be|}(1+|x|)^{-\ka-|\be|}\,,
\end{equation}
so we find
\begin{equation}\label{phih}
    \|\phi\|_{\ka+l,\,l}
    \leq\con\,\|H_\ka\phi\|_{\ka+\ep+l,\,l}\,.
\end{equation}
Furthermore, define
\begin{equation}\label{phias}
    \phi_\as(x)=\int_0^\infty u^{\ka-1}\psi(ux)du\,,
\end{equation}
which is homogeneous of degree $-\ka$ and $\C^\infty$ outside
$x=0$. Then
\begin{equation}\label{phiphias}
    D^\beta[\phi-\phi_\as](x)
    =-\int_1^\infty u^{\ka+|\beta|-1}[D^\beta\psi](ux)\,du\,.
\end{equation}
Estimating the integral one finds that for $|x|\geq1$ we have
\begin{equation}\label{estrest}
    |x|^{\ka+\ep+|\be|}\big|D^\beta[\phi-\phi_\as](x)\big|
\leq\con\|H_\ka\phi\|_{\ka+\ep+|\beta|,\,|\beta|}\,.
\end{equation}
It follows that
\begin{equation}\label{as}
    \phi_\as(x)=\lim_{R\to\infty}R^\ka\phi(Rx)\,,
\end{equation}
and this asymptote is independent of the choice of the central
point $O$. With the use of the estimates (\ref{phih}) and
(\ref{estrest}) one finds that in the special case of vanishing
asymptote we have a~bound stronger than (\ref{phih}):
\begin{equation}\label{phihstr}
    \|\phi\|_{\ka+\ep+l,\,l}
    \leq\con\,\|H_\ka\phi\|_{\ka+\ep+l,\,l}\quad
    \text{iff}\quad \phi_\as=0\,.
\end{equation}

The estimates (\ref{hphi}) and (\ref{phih}) imply that
\begin{equation}\label{hs}
    H_\ka\Sc_{\ka+\ep}\subset\Sc_{\ka+\ep}
    \subset H_\ka\Sc_\ka\,,
\end{equation}
so $H_\ka^{-1}$ maps $\Sc_{\ka+\ep}$ bijectively onto
$\Sc^\ka_{\ka+\ep}=H_\ka^{-1}(\Sc_{\ka+\ep})$, and $\Sc_{\ka+}$
onto
\begin{equation}\label{sphom}
    \Sc^\ka_{\ka+} =\bigcup_{0<\ep<1}\Sc^\ka_{\ka+\ep}\,.
\end{equation}
We also use $H_\ka^{-1}$ to transfer the topological inductive
limit structure from $\Sc_{\ka+}$ to $\Sc^\ka_{\ka+}$. Thus the
topology $\T^\ka_{\ka+}$ of $\Sc^\ka_{\ka+}$ is the inductive
limit of topologies $\T^\ka_{\ka+\ep}$ of $\Sc^\ka_{\ka+\ep}$,
which are determined by the seminorms
$\|\phi\|^\ka_{\ka+\ep+l,\,l}=\|H_\ka\phi\|_{\ka+\ep+l,\,l}$,
$l=0,1,\ldots$. From (\ref{hs}) and the relations between
seminorms (\ref{hphi}), (\ref{phih}) and (\ref{phihstr}) we have
\begin{equation}\label{sss}
    \Sc_{\ka+}\subset\Sc^\ka_{\ka+}\subset\Sc_\ka\,,
    \quad \T_{\ka+}\sim\T^\ka_{\ka+}>\T_\ka\,;
\end{equation}
here the sign $\sim$ means that the topology to the left is equal
to the topology induced by the one to the right. Space
$\Sc_{\ka+}$ consists of all those functions $\phi$ in
$\Sc^\ka_{\ka+}$ for which $\phi_\as=0$, and forms a~closed
subspace of $\Sc^\ka_{\ka+}$ (if $\phi_\la\to\phi$ in
$\Sc_{\ka+}^\ka$, then this limit is also achieved in the topology
$\T_\ka$, which implies that $\phi_\as=0$ if $\phi_{\la\as}=0$).

The topological inductive limit structure summarized in
(\ref{sphom}) and (\ref{sss}) is independent of the choice of
reference system $(O,(e_i))$. This is immediate for the change of
basis. Also, the independence of $\Sc^\ka_{\ka+\ep}$ and
$\Sc^\ka_{\ka+}$ as sets from the choice of $O$ follows
immediately from the independence of $\Sc_{\ka+\ep}$ from that
choice. It remains to be shown that $\T^\ka_{\ka+\ep}$ is
translationally invariant. This follows from the estimates
\begin{equation}\label{trinv}
    \|H_\ka\phi_a\|_{\ka+\ep+l,\,l}\leq
    \con\,\|H_\ka\phi\|_{\ka+\ep+l,\,l}+
    \con\,\|H_\ka\phi\|_{\ka+\ep+l+1,\,l+1}\,,
\end{equation}
which are obtained by writing
$H_\ka\phi_a(x)=[H_\ka\phi]_a(x)+a\cdot\p\phi_a(x)$ and using
(\ref{transl}) and (\ref{phih}).

\subsection{Spaces
$\boldsymbol{\sloc(\Sc^\ka_{\ka+})}$}\label{test-loc}

Let $\Oc$ be any open set in $\M$ with the closure $\ov{\Oc}$. The
set of fields in the space $\Sc^\ka_{\ka+}$ with support in
$\ov{\Oc}$ forms a~closed subspace equipped with the induced
topology; we denote it $\Sc^\ka_{\ka+}(\ov{\Oc})$. Let $V_\pm$ be
the open future (past) lightcones in $M$. We shall use notation
$\C_\pm$ for any of the open sets $\C_\pm=P_\pm+V_\pm$,
$P_\pm\in\M$, and we shall also write $\C=\C_+\cup\C_-$ for any
$\C_\pm$ such that $\C_+\cap\C_-\neq\emptyset$. The family
$\Sc^\ka_{\ka+}(\ov{\C})$ with the induced topologies
$\T^\ka_{\ka+}(\ov{\C})$ forms an increasing net of locally convex
spaces. It~is now easy to see that the sum (s.l. stands for
``spatially local'')
\begin{equation}\label{il}
    \sloc(\Sc^\ka_{\ka+})
    =\bigcup_\C\Sc^\ka_{\ka+}(\ov{\C})
    \subset\Sc^\ka_{\ka+}
\end{equation}
forms a~strict inductive limit and the limit topology
$\sloc(\T^\ka_{\ka+})$ satisfies
\begin{equation}\label{ilt}
    \T^\ka_{\ka+}(\ov{\C})
    \sim\sloc(\T^\ka_{\ka+})
    >\T^\ka_{\ka+}\,.
\end{equation}
For each $\phi$ in this space there is: $\supp\phi_\as\subseteq
\ov{V_+\cup V_-}$.

Finally, we note that the derivative $\p_i$ maps continuously
$\Sc_\ka\mapsto\Sc_{\ka+1}$ and
$\Sc^\ka_{\ka+\ep}\mapsto\Sc^{\ka+1}_{\ka+1+\ep}$, while
multiplication by $x^i$ maps continuously
$\Sc_{\ka+1}\mapsto\Sc_\ka$ and
\mbox{$\Sc^{\ka+1}_{\ka+1+\ep}\mapsto\Sc^\ka_{\ka+\ep}$}. Then
similar continuous connections also take place between pairs of
spaces of the type $\Sc_{\ka+}$, $\Sc^\ka_{\ka+}$,
$\Sc^\ka_{\ka+}(\ov{\C})$ and $\sloc(\Sc^\ka_{\ka+})$.

\subsection{Spaces $\boldsymbol{\J_\inc}$ and
$\boldsymbol{\J_\out}$}

Choose now $\ka=3$ and consider the space $\sloc(\Sc^3_{3+})$ of
vector fields. We denote by $\J_\inc$ ($\J_\out$) the subspace of
fields $J$ which satisfy the following additional conditions:
\begin{equation}\label{V}
    \p\cdot J=0\,,\quad x\wedge J_\as=0\,,\quad
    \supp J_\as\subseteq\ov{V_-}\  (\text{resp.}\ \ov{V_+})   \,,
\end{equation}
(compare (\ref{falloff}) and (\ref{elas})). Let $J_\lambda$ be any
net in $\J_\inc$ ($\J_\out$), $J$ an element of
$\sloc(\Sc^3_{3+})$, and $J_\lambda\to J$. The mapping
$J\mapsto\p\cdot J$ is continuous (between suitable spaces -- see
the end of the last subsection) so the first condition is
conserved under the limit and $\p\cdot J=0$. As
$\sloc(\T^\ka_{\ka+})$ is stronger than the topology induced by
$\T_\ka$ it is easy to see that the support properties of
$(J_\lambda)_\as$ are conserved under the limit, so $J$ satisfies
the third condition. Finally, using the continuity of the mapping
$J\mapsto x\wedge J$ and again the conservation of support
properties one finds that $J$ satisfies the second condition. Thus
$\J_\inc$ and $\J_\out$ are closed subspaces of
$\sloc(\Sc^3_{3+})$. We shall write $\J_\as$ for $\J_\inc$ or
$\J_\out$, and we shall also set $J_\inc=J_\as$ in $\J_\inc$, and
$J_\out=J_\as$ in $\J_\out$. We denote by $\J_\as(\ov{\C})$ the
subspace of $\J_\as$ consisting of currents supported in
$\ov{\C}$.

Let $s\in\mR$ and $l$ be a~future-pointing lightlike vector and
choose a~region $\C$. It is shown by a~straightforward calculation
that for $\ka>2$
\begin{equation}\label{est}
    \int_\C\frac{\delta(s-x\cdot l)}{(|x|+1)^\ka}dx
    \leq\frac{\con(\C)}{(|s|+1)^{\ka-3}}\,,
\end{equation}
where $\delta(.)$ is the Dirac delta function. If vectors $l$ are
scaled to $l^0=1$ then the bounding constant in the above relation
is $l$-independent. Therefore for $J\in\J_\as$ the integral
\begin{equation}\label{Vsl}
    V(s,l)=\int J(x)\delta(s-x\cdot l)\,dx
\end{equation}
is absolutely convergent, the function $V(s,l)$ is homogeneous of
degree $-1$: $V(\mu s,\mu l)=\mu^{-1}V(s,l)$ for $\mu>0$, and if
$l$'s are scaled to $l^0=1$ then it is bounded. We denote
$L_{ab}=l_a\p/\p l^b-l_b\p/\p l^a$, \mbox{$X_{ab}=x_a\p/\p
x^b-x_b\p/\p x^a$} and observe that
\begin{equation}\label{dd}
    (L_{ab}+X_{ab})\delta(s-x\cdot l)=0\,,\quad
    (s\p_s+x\cdot\p+1)\delta(s-x\cdot l)=0\,.
\end{equation}
Using these identities we find that $V(s,l)$ is infinitely
differentiable (outside the vertex of the cone; operators $L_{ab}$
incorporate all intrinsic derivatives in the cone), and we have
(dot denotes the derivative $\p/\p_s$)
\begin{equation}\label{vdot}
    sL_{a_1b_1}\ldots L_{a_nb_n}\dV_c(s,l)=
    \int \delta(s-x\cdot l)X_{a_1b_1}\ldots
    X_{a_nb_n}H_3J_c(x)\,dx\,.
\end{equation}
Let $J$ be in $\J_\as\cap\Sc^3_{3+\ep}(\ov{\C})$. Then estimating
the integrand by
\begin{equation}\label{xhj}
    |X_{a_1b_1}\ldots X_{a_nb_n}H_3J_c(x)|\leq
    \con\sum_{l=0}^n\|H_3J\|_{3+\ep+l,\,l}(1+|x|)^{-3-\ep}
\end{equation}
and using (\ref{est}) we obtain
\begin{equation}\label{estv}
    |L_{a_1b_1}\ldots L_{a_nb_n}\dV(s,l)|\leq
    \con(\C)\sum_{l=0}^n\|J\|^3_{3+\ep+l,\,l}(1+|s|)^{-1-\ep}
\end{equation}
for $l$'s scaled to $l^0=1$. The limits $V(\pm\infty,l)$ are
determined by $J_\as$ and one finds
\begin{equation}\label{limv}
 \begin{split}
    V(-\infty,l)=\int J_\inc(x)\delta(x\cdot l+1)\,dx\,,\quad
    V(+\infty,l)=0\qquad \text{for}\ J\in\J_\inc\,,\\
    V(-\infty,l)=0\,,\quad
    V(+\infty,l)=\int J_\out(x)\delta(x\cdot l-1)\,dx
    \qquad \text{for}\ J\in\J_\out\,.
 \end{split}
\end{equation}
In addition the following identities are satisfied
\begin{equation}\label{llimv}
    L_{[ab}V_{c]}(\pm\infty,l)=0\,,\quad l\cdot V(s,l)=0\,.
\end{equation}
The first of them is the consequence of the second condition in
(\ref{V}). To prove the second one we observe that
$\p_s(s\,l\cdot\dV(s,l))=\int\delta(s-x\cdot l)H_4\p\cdot
J(x)\,dx=0$ by (\ref{vdot}) and conservation of $J$, and then the
result follows by (\ref{limv}). For currents with non-vanishing
both future and past asymptotes $J_\as$ the rhs of the second
equation in (\ref{llimv}) is the total charge.

The radiation potential produced by the current $J\in\J_\as$ is
completely determined by $\dV(s,l)$ according to the formula
\cite{her98}
\begin{equation}\label{JA}
    A(x)=-\frac{1}{2\pi}\int\dV(x\cdot l,l)\,d^2l\,,
\end{equation}
which follows from the representation
 $D(x)=-(1/8\pi^2)\int\delta'(x\cdot l)\,d^2l$. Here $d^2l$ is the
invariant measure on the set of null directions: we remind the
reader that if $f(l)$ is homogeneous of degree $-2$ then the
integral
\begin{equation}\label{d2l}
    \int f(l)\,d^2l=\int f(1,\vec{l})\,d\Omega(\vec{l})\,,
\end{equation}
where $d\Omega(\vec{l})$ is the solid angle measure in the
direction of the unit 3-vector $\vec{l}$, is independent of the
choice of Minkowski basis, and satisfies
\begin{equation}\label{parts}
    \int L_{ab}f(l)\,d^2l=0\,.
\end{equation}
Using (\ref{JA}) to express $A_1$ and $A_2$ in the symplectic form
(\ref{sfj}) we find that the integrand in that form is absolutely
integrable and one obtains
\begin{equation}\label{sfv}
    \{J_1,J_2\}=\frac{1}{4\pi}
    \int(\dV_1\cdot V_2-\dV_2\cdot V_1)(s,l)\,ds\,d^2l\,,
\end{equation}
so $\J_\as$ becomes a~symplectic space. Moreover, using
(\ref{estv}) and the fact that one of the asymptotic limits
$V(\pm\infty,l)$ vanishes, one easily obtains the estimate
\begin{equation}\label{sfest}
    \big|\{J_1,J_2\}\big|\leq\con(\C)\,
    \|J_1\|^3_{3+\ep,\,0}\,\|J_2\|^3_{3+\ep,\,0}\,.
\end{equation}
Using properties of $V_i$ it is easy to find the kernel of the
symplectic form:
\begin{equation}\label{ksf}
    \Ker\{.\,,.\}=\{J\in\J_\as\mid l\wedge V=0\}\,.
\end{equation}
If $J_v$ is a~current of a~point particle carrying charge $e$ and
moving freely along any world-line parallel to the four-velocity
$v=p/m$ then it is easy to find that
\begin{equation}\label{sypo}
    \{J,J_{p/m}\}=\frac{e}{4\pi}
    \int\frac{p\cdot\Delta V(l)}{p\cdot l}d^2l\,,
\end{equation}
where
\begin{equation}\label{delv}
    \Delta V(l)=V(+\infty,l)-V(-\infty,l)=\mp V(\mp\infty,l)
\end{equation}
in the ``in'' and ``out'' case respectively. By the substitution
$p\to w=p+iq$ the rhs of (\ref{sypo}) may be analytically extended
to the domain $\{w\mid p^2>0\ \text{or}\ q^2>0\}$; we denote this
extension $F_J(w)$. It is shown in Appendix A that there exist
continuous limit functions on $M$:
$F_{J\pm}(p)=\lim_{\la\searrow0}F_J(p\pm\la iq)$, where $q$ is a
timelike future-pointing vector, with $F_{J\pm}(p)=F_J(p)$ for
$p^2>0$.

The electromagnetic potential represented by (\ref{JA}) is
infrared singular (has a~spacelike tail of the decay rate of the
Coulomb field) if $\Delta V(l)\neq0$, and this function
characterizes this singularity completely. We observe that our
form of the symplectic structure remains well-defined for those
fields. Also, note that (\ref{sypo}) vanishes for infrared-regular
fields, so $W(J)$ for such currents commutes with charged particle
field (see (\ref{wepsi},\ref{s}) and below).

\subsection{Space $\boldsymbol{\Di}$}

For the space $\C^\infty(\M)$ of fields of given geometric type we
define subspaces
\begin{equation}\label{sbox}
    \Sb_\ka=\{\phi\in\C^\infty\mid\sup_x
    (1+|x|)^{\ka+n}|\Box^nD^\beta\phi_j(x)|<\infty\,,\
    \forall n=0,1,\ldots\,,\ \forall \beta\,,\ \forall j\}
\end{equation}
(independent of the choice of reference system). We also introduce
subspaces $\Sb_\ka(\ov{\C})$ of functions with support in
$\ov{\C}$ and the algebraic inductive limit space
\begin{equation}\label{slbox}
    \sloc(\Sb_\ka)=\bigcup_{\C}\Sb_\ka(\ov{\C})\,.
\end{equation}

Consider the space of $4$-spinor fields of the type
$\Di=\sloc(\Sb_5)$; we shall also write
$\Di(\ov{\C})=\Sb_5(\ov{\C})$. The Fourier transforms of fields in
that space (after fixing the origin) are among continuous
functions vanishing faster then polynomially at infinity. Two test
fields $\chi_1$ and $\chi_2$ are in one class producing the same
Dirac field according to (\ref{psch}) if, and only if, the
restrictions of $\wh{\chi_1}$ and $\wh{\chi_2}$ to the hyperboloid
$p^2=m^2$ are equal. For $\chi\in\Di$ we shall denote by $[\chi]$
the class of fields producing the same Dirac field as $\chi$, and
by $[\Di]$ the quotient space of these classes; also,
$[\Di(\ov{\C})]$ will denote the set of classes $[\chi]$ with
$\chi\in\Di(\ov{\C})$.

Let $\chi\in\Sb_5(\ov{\C})$ and $k>5$. Then we have
\begin{equation}\label{5n}
 \begin{split}
    &\chi=\chi_k+(m^2+\Box)\chi'_k\,,\quad
    \chi_k=\Big(\frac{-\Box}{m^2}\Big)^{k-5}\chi\in\Sb_k(\ov{\C})\,,\\
    &\chi'_k=\frac{1}{m^2}\bigg[1+\Big(\frac{-\Box}{m^2}\Big)+\ldots
    +\Big(\frac{-\Box}{m^2}\Big)^{k-6}\bigg]\chi\in\Sb_5(\ov{\C})\,.
 \end{split}
\end{equation}
Therefore for each $\chi\in\Di$ the class $[\chi]$ contains for
each $k\geq5$ a~field $\chi_k\in\Sb_k$ with the same support
properties as $\chi$.

For each $J\in\J_\as$ we define a~linear operator $S_J$ in $[\Di]$
as follows. For \mbox{$\chi\in\Sb_5(\ov{\C})$} with
$\C=\C_+\cup\C_-$ we find any $\chi_k\in[\chi]\cap\Sb_k(\ov{\C})$
with $k\geq10$ and split $\chi_k=\chi_{k+}+\chi_{k-}$,
$\chi_{k\pm}\in\Sb_k(\ov{\C_\pm})$, which is possible, as shown in
Appendix \ref{phase}. We put
\begin{equation}\label{sj}
    S_J[\chi]=[\chi']\,,\quad\text{where}\quad
    \wh{\chi'}(p)=e^{-iF_{J+}(p)}\wh{\chi_{k+}}(p)
    +e^{-iF_{J-}(p)}\wh{\chi_{k-}}(p)\,,
\end{equation}
and the functions $F_{J\pm}$ were defined at the end of the last
subsection. The results of Appendix \ref{phase} guarantee that
this is a~correct definition, i.e. the function $\chi'$ is in
$\Sb_5(\ov{\C})$ and the class $[\chi']$ is independent of the
choice of $\chi_k$ and its split into $\chi_{k\pm}$. Moreover, it
is easy to convince oneself that
\begin{equation}\label{sjb}
    S_{J_1}\,S_{J_2}=S_{J_1+J_2}\,,\quad S_0=\id\,,\quad
    S_J[\Di(\ov{\C})]=[\Di(\ov{\C})]\,,
\end{equation}
so $S_J$ is bijective for each $J$ and the mapping
$S_\bullet:J\mapsto S_J$ is a~homomorphism of the additive group
of currents into the group of automorphisms of $[\Di]$ and of each
of the spaces $[\Di(\ov{\C})]$. The kernel of this homomorphism is
the subgroup of currents given by
\begin{equation}\label{ksj}
    \Ker S_\bullet=\{J\in\J_\as\mid \{J,J_v\}=2k\pi,\
    k=0,\pm1,\ldots\,\}\,.
\end{equation}
The mappings are unitary with respect to the scalar product
(\ref{preh}):
\begin{equation}\label{sjuni}
    \text{if}\quad[\chi'_i]=S_J[\chi_i]\quad\text{then}\quad
    \<\chi'_1,\chi'_2\>=\<\chi_1,\chi_2\>\,.
\end{equation}

\newpage

\setcounter{equation}{0}

\section{Asymptotic algebras of fields}\label{alg}

The structures of the last section alow now for a~rigourous
formulation of our ideas in the algebraic form.

\subsection{$\boldsymbol{}^*$-algebras $\boldsymbol\B_\mathbf{as}$}

We define the $^*$-algebras of fields. For each $J\in\J_\as$ we
assume an element of the algebra $W_\as(J)$, and for each
$\chi\in\Di$ an element $\psi_\as(\chi)$. We also assume a~unit
element $E$ and impose the algebraic relations
\begin{equation}\label{stalg}
 \begin{split}
    W_\as(J)^*&=W_\as(-J)\,,\quad W_\as(0)=E\,,\\
    W_\as(J_1)W_\as(J_2)&=\exp[-\tfrac{i}{2}\{J_1,J_2\}]W_\as(J_1+J_2)\,,\\
    [\psi_\as(\chi_1),\psi_\as(\chi_2)]_+&=0\,,\qquad
    [\psi_\as(\chi_1),\psi_\as(\chi_2)^*]_+=\<\chi_1,\chi_2\>\,,\\
    W_\as(J)\psi_\as(\chi)&=\psi_\as(\chi')W_\as(J)\,,\quad\text{where}
    \quad [\chi']=S_J[\chi]\,.
 \end{split}
\end{equation}
We want to identify those elements $W_\as(J)$ which generate the
same relations. The following is a~subgroup of the additive group
of currents $\J_\as$:
\begin{multline}\label{ksfsj}
    \J_\as^0=\Ker\{.\,,.\}\cap\Ker S_\bullet\\
    =\Big\{J\in \J_\as\mid V(s,l)=l\al(s,l),\
    (e/4\pi)\int\Delta\al(l)\,d^2l=2k\pi,\ k=0,\pm1,\ldots\,\Big\}\,,
\end{multline}
where
$\Delta\al(l)=\al(+\infty,l)-\al(-\infty,l)=\mp\al(\mp\infty,l)$
in the ``in'' and ``out'' case respectively. We shall denote by
$[\J_\as]$ the quotient group $\J_\as/\J_\as^0$, and set
$W_\as(J_1)=W_\as(J_2)$ if $[J_2]=[J_1]$. Similarly we identify
$\psi_\as(\chi_1)=\psi_\as(\chi_2)$ if $[\chi_1]=[\chi_2]$. After
these identifications we shall call the $^*$-algebra generated by
the relations (\ref{stalg}) the field $^*$-algebra $\B_\as$. For
each $\C$ the elements $W_\as(J)$ and $\psi_\as(\chi)$ with
$J\in\J_\as(\ov{\C})$ and $\chi\in\Di(\ov{\C})$ generate a
subalgebra, denoted $\B_\as(\ov{\C})$, and we have
$\B_\as=\cup_{\C}\B_\as(\ov{\C})$.

This construction can also be characterized as follows. The
relations (\ref{stalg}) generate a~$^*$-algebra. The elements
$A\big(W_\as(J)-E\big)$, $J\in[0]$, and $A\psi_\as(\chi)$,
$\chi\in[0]$, where $A$ goes over all elements of the algebra,
generate a~two-sided ideal of the algebra. The quotient of the
algebra through this ideal is the $^*$-algebra $\B_\as$, as=in or
out.

The elements $\psi_\as(\chi)$ generate a~subalgebra $\B_\as^+$ of
the CAR type, and elements $W_\as(J)$ -- a~subalgebra $\B_\as^-$
of the CCR type. Each element of $\B_\as$ may be brought to the
form $\sum_{i=1}^k C_iW_\as(J_i)$, where $C_i\in\B_\as^+$ and with
currents $J_i$ such that $[J_i]\neq[J_j]$ for $i\neq j$. The last
relation in (\ref{stalg}) may be used to define a~group of
automorphisms of $\B_\as^+$:
\begin{equation}\label{beta}
 \begin{split}
    &\be_J(C)=W_\as(J)CW_\as(-J)\,,\quad C\in\B_\as^+\,,\\
    &\be_{J_1}\be_{J_2}=\be_{J_1+J_2}\,,\quad \be_0=\id\,.
 \end{split}
\end{equation}

The universal covering group $\Pg$ of the Poincar\'e group has a
representation in the automorphism group of algebra $\B_\as$.
After choosing the origin in $\M$ each element in $\Pg$ is
represented by $(a,A)$, $a\in M$, $A\in SL(2,\mC)$, and the
respective automorphism $\al_{a,A}$ is given by the standard
formulas
\begin{equation}\label{poin}
 \begin{split}
    \al_{a,A}\big(W_\as(J)\big)=W_\as(T_{a,A}J)\,,&\quad
    [T_{a,A}J](x)=\Lambda(A)J(\Lambda(A)^{-1}(x-a))\,,\\
    \al_{a,A}\big(\psi_\as(\chi)\big)=\psi_\as(R_{a,A}\chi)\,,&\quad
    [R_{a,A}\chi](x)=S(A)\chi(\Lambda(A)^{-1}(x-a))\,,
 \end{split}
\end{equation}
where $\Lambda(A)$ and $S(A)$ are elements of the vector and
$4$-spinor representations of $SL(2,\mC)$ respectively.

In the above construction of the algebras $\B_\as$ we identified
elements labelled by test functions falling into a~common
equivalence class. The net effect of these identifications is that
the elements $\psi(\chi)$ could be labelled by the restriction of
the Fourier transform of $\chi$ to the hyperboloid $p^2=m^2$, and
the elements $W(J)$ -- by the corresponding classes of $V$'s up to
addition of an element of $\J_\as^0$. If formulated in this way,
the algebra $\B_\out$ is a~subalgebra of the algebra $\B$ of Ref.\
\cite{her98}, Eqs.\ (3.40-43). The use of the present test spaces
adds to the elements of the algebra the spacetime localization
properties. In the following (sub)sections we use results of Ref.\
\cite{her98}, although with some modifications indicated below.
The change from $\B_\out$ to $\B_\inc$ is almost trivial in all
what follows and we shall not separate these two cases.

To be precise, there is one minor difference in the use of $V$'s
in Ref.\ \cite{her98} and the present paper. Let us concentrate on
the ``out'' case. Here we have $V(-\infty,l)=0$ in that case,
while in \cite{her98} we used variable $V^\out(s,l)$ for which
$V^\out(+\infty,l)=0$. The relation between the two variables is
$V^\out(s,l)=V(s,l)-V(+\infty,l)$. It is easy to check that the
value of the symplectic form remains unchanged under this
transformation, and also $\Delta V^\out(l)=\Delta V(l)$, so $S_J$
is unaffected. The convention of $V$ is more naturally connected
with the outgoing test current, while the convention of $V^\out$
is connected with the description of the future null asymptotics
of the radiation potential (\ref{JA}): for a~lightlike
future-pointing vector $k$ one has
\begin{equation}\label{las}
    \lim_{R\to\infty} RA(x+Rk)=V^\out(x\cdot k,k)\,.
\end{equation}
Similar connections, with past interchanged with future, take
place in the ``in'' case.

\subsection{Selection of representations}

We consider $^*$-representations $\pi$ of the algebra $\B_\as$ by
bounded operators in a~Hilbert space $\Hc$ (all representations
considered will be $^*$-representations and we suppress this
qualification in the sequel). We are interested only in regular
representations -- those for which all one-parameter groups
$\la\mapsto \pi\big(W_\as(\la J)\big)$ are strongly continuous;
the Weyl exponentiation is a~technical device, which should be
invertible on the level of representations having physical
significance. Moreover, we further restrict attention to the
translationally covariant representations with positive energy.
This means that there exists a~representation of the translation
group by unitary operators $U(a)$ in $\Hc$ which implement the
automorphisms $\al_a\equiv\al_{a,\1}$, i.e.\
$\pi\big(\al_a(B)\big)=U(a)\pi(B)U^*(a)$ for each $B\in\B_\as$,
and whose spectrum is contained in~$\ov{V_+}$.

Let $\pi_F$ be the standard positive energy Fock representation of
$\B_\as^+$ on the Hilbert space $\Hc_F$, with the Fock vacuum
vector denoted $\W_F$, and $\pi_r$ be a~regular, translationally
covariant positive energy representation of $\B_\as^-$ on $\Hc_r$.
Define the following operators $\pi(A)$ on the space
$\Hc=\Hc_F\otimes\Hc_r$ by
\begin{equation}\label{rep}
 \begin{split}
    \pi(C)&=\pi_F(C)\otimes\id_r\,,\quad C\in\B_\as^+\,,\\
    \pi(W_\as(J))[\pi_F(B)\W_F\otimes\varphi]&=
    \pi_F(\be_JB)\W_F\otimes\pi_r(W_\as(J))\varphi\,,\quad B\in\B_\as^+\,.
 \end{split}
\end{equation}
Then $\pi$ extends to a~regular, translationally covariant
positive energy representation of $\B_\as$. Conversely, if $\pi$
is a~representation of $\B_\as$ with these properties then up to a
unitary equivalence it has the form given by (\ref{rep}).

The theorem formulated in the last paragraph results from a~slight
modification of the Theorem 4.4 of Ref.\ \cite{her98}, taken over
here by the remarks of the two last paragraphs of the preceding
subsection. The modification is twofold. First, the theorem is
formulated in terms of a~$C^*$-algebra $\F$ generated by $\B$, but
in the proofs of this theorem and its lemmas one can replace $\F$
by $\B$. Second, we need to replace $\B$ by $\B_\as$. This however
poses no problem, one only has to use the fact that our present
test space $\Di$ is dense in the Hilbert space of the scalar
product (\ref{preh}).

We now restrict the choice of representations still further. We
demand that the elements $\pi(W_\as(J))$ are gauge-invariant when
acting on the subspace \mbox{$\W_F\otimes\Hc_r$}. The physical
motivation for that seems plausible enough -- the electromagnetic
field alone should be gauge-invariant. Let $[[\J_\as]]$ be the
quotient space $\J_\as/\Ker\{.\,,.\}$ with elements denoted by
$[[J]]$. The assumed gauge-invariance means that \linebreak[4]
\mbox{$\pi_r(W_\as(J_1))=\pi_r(W_\as(J_2))$} if $[[J_1]]=[[J_2]]$.
This is equivalent to the assumption that $\pi_r$ is a
representation of the algebra $[[\B_\as^-]]$ (in which such
elements $W_\as(J_1)$ and $W_\as(J_2)$ have been identified).

\subsection{$\boldsymbol C^*$ field algebras $\boldsymbol\F_\mathbf{as}$}

We can now equip the algebra $\B_\as$ with a~$C^*$-norm defined by
$\|A\|=\|\pi(A)\|$, where $\pi$ is any representation in the class
defined by the above assumptions. This norm is independent of the
choice of the particular representation $\pi$. The uniqueness
follows in three steps. First, the representation $\pi_F$ extends
uniquely to the Fock representation of the full (unique) CAR Dirac
fields algebra -- this is because fields $[\chi]$, $\chi\in\Di$,
are dense in the Hilbert space defined by the product (\ref{preh})
(as, in particular, compactly supported $\chi$'s are).  Second, as
the symplectic form is nondegenerate on $[[\J_\as]]$, the algebra
$[[\B_\as^-]]$ generates the unique $C^*$ Weyl algebra. Each
representation $\pi_r$ in the assumed class extends to a
representation of this Weyl algebra. Third, these extended
representations $\pi_F$ and $\pi_r$ are faithful, so the operator
norm on $\pi_F(\B_\as^+)\otimes_\mathrm{alg}\pi_r(\B_\as^-)$ is
independent of the choice of $\pi_r$. This is sufficient to
conclude the claimed uniqueness.

We shall call the $C^*$-algebra generated by $\B_\as$ equipped
with the norm introduced above the $C^*$ asymptotic field algebra,
$\F_\as=\ov{\B_\as}^{\scriptscriptstyle\|.\|}$. Each
representation $\pi$ of $\B_\as$ in the assumed class extends to a
faithful representation of $\F_\as$. We also introduce algebras
$\F_\as(\ov{\C})=\ov{\B_\as(\ov{\C})}^{\scriptscriptstyle\|.\|}$.

We note that the present construction of the algebras $\F_\as$ is
more restrictive than the one leading from $\B$ to $\F$ in Ref.\
\cite{her98}. There the construction of $\F$ was based on all
possible Hilbert space representations of $\B$. However, now I
think that this is both more involved and unjustified. Some of the
elements of this larger algebra could be brought to zero in
representations having physical interpretation.

\subsection{Examples of the representations
$\boldsymbol\pi_r$}\label{alg-ex}

Let $\rho$ be a~real smooth function on $\M$ of compact support,
such that $\int\rho(x)dx=1$. For each $J\in\J_\as(\ov{\C})$ we
denote
\begin{equation}
 J_\rho=J-\rho*J_\as\,,\quad
 (\rho*J_\as)(x)=\int \rho(x-y)J_\as(y)\,dy\,.
\end{equation}
Modifying slightly the steps of Appendix \ref{inout} one finds
that $\rho*J_\as, J_\rho\in\J_\as(\ov{\C})$.\linebreak The
asymptote of $\rho*J_\as$ is equal to $J_\as$, therefore the
asymptote of $J_\rho$ vanishes, $J_{\rho\as}=0$. Moreover, taking
into account the support property of $J_\as$ one finds
\begin{multline*}\label{}
    V(s,l)-V_\rho(s,l)=\int\delta(s-x\cdot l)(\rho*J_\as)(x)dx\\
    =\int\rho(z)\bigg\{\int
    \delta(s-z\cdot l-y\cdot l)J_\as(y)\,dy\bigg\}\,dz\\
    =\int\theta(\mp(s-z\cdot l))\rho(z)dz
    \int\delta(y\cdot l\pm1)J_\as(y)dy\,,
\end{multline*}
where the upper and lower signs refer to the ``in'' and ``out''
case respectively and the last step results from rescaling
$s-z\cdot l$ in the inner integral to $\mp1$ respectively. If we
introduce
\begin{equation}\label{Hf}
    H(s,l)=\int\sgn(s-z\cdot l)\rho(z)dz\,,\quad
    \text{with}\quad \lim_{s\to\pm\infty}H(s,l)=\pm1\,,
\end{equation}
then
\begin{equation}\label{vvf}
    V(s,l)=V_\rho(s,l)+\tfrac{1}{2}[1\mp H(s,l)]\,V(\mp\infty,l)\,.
\end{equation}
Using this split in (\ref{sfv}) it is easy to show that
\begin{equation}\label{sfvsplit}
    \{J_1,J_2\}=\{J_{1\rho},J_{2\rho}\}+\{J_1,\rho*J_{2\as}\}
    -\{J_2,\rho*J_{1\as}\}\,.
\end{equation}
Currents $J_{i\rho}$ produce infrared-regular fields and the first
term on the rhs coincides with the standard symplectic form for
such fields, and expressed in terms of $V$'s reads
\begin{equation}\label{sfvreg}
    \{J_{1\rho},J_{2\rho}\}=\frac{1}{4\pi}
    \int(\dV_{1\rho}\cdot V_{2\rho}
    -\dV_{2\rho}\cdot V_{1\rho})(s,l)\,ds\,d^2l\,,
\end{equation}
The rest of the rhs of (\ref{sfvsplit}) is another symplectic form
which we now transform. By a~straightforward calculation we have
\begin{equation}\label{sfvrho}
    \{J_1,\rho*J_{2\as}\}
    =\int\bigg(\frac{1}{4\pi}\int\dV_1H(s,l)ds\bigg)\cdot
    \Delta V_2(l)\,d^2l\,,
\end{equation}
with $\Delta V(l)$ defined in (\ref{delv}). We recall from Ref.\
\cite{her98} that for vector fields $f(l)$ homogeneous of degree
$-1$ and orthogonal to $l$ the scalar product $(f,g)_0=-\int
f(l)\cdot g(l)\,d^2l$ defines a~Hilbert space $\Hc_0$ of
equivalence classes, and that the fields satisfying in addition
$L\wedge f=0$ (cf.\ Eq.\ (\ref{llimv})) span its subspace
$\Hc_{IR}$. (All fields differing by fields of the form $l\al(l)$
fall into one class. To simplify notation we suppress the square
brackets which were used in \cite{her98} to distinguish a~class
$[f]$ from the field $f$.) Following the notation of \cite{her98}
we denote $p(\dV)=\frac{1}{2\pi}\Delta V$. Also, we set
$h=\pi\dot{H}$, and again following \cite{her98} write $r_h(\dV)$
for the orthogonal projection in $\Hc_0$ onto $\Hc_{IR}$ of
$\frac{1}{2}\int\dV H(s,l)ds$. Now we can write
\begin{equation}\label{sfw}
    \{J_1,\rho*J_{2\as}\} -\{J_2,\rho*J_{1\as}\}
    =\{p(\dV_1)\oplus r_h(\dV_1),p(\dV_2)\oplus
    r_h(\dV_2)\}_{IR}\,,
\end{equation}
where
\begin{equation}\label{sfir}
    \{g_1\oplus k_1,g_2\oplus k_2\}_{IR}=
    (g_1,k_2)_{\Hc_{IR}}-(g_2,k_1)_{\Hc_{IR}}
\end{equation}
is a~nondegenerate symplectic form on the space
$\Hc_{IR}\oplus\Hc_{IR}$.

The additive split of the symplectic form into the parts
(\ref{sfvreg}) and (\ref{sfw}) suggests the following
construction. Let $\hat{W}(V)$ generate the standard Weyl algebra
over the space of infrared-regular fields with the symplectic form
(\ref{sfvreg}), and \mbox{$w(g\oplus k)$} generate the Weyl
algebra over the symplectic space given by (\ref{sfir}). Choose
representations $\pi_\mathrm{reg}$ and $\pi_\mathrm{sing}$ of
these two algebras. Then the formula
\begin{equation}\label{pir}
    \pi_r(W(J))=\pi_\mathrm{reg}\big(\hat{W}(V_\rho)\big)\otimes
 \pi_\mathrm{sing}\big(w(p(\dV)\oplus r_h(\dV))\big)
\end{equation}
defines a~representation of the algebra $\B_\as^-$. If
$\pi_\mathrm{reg}$ and $\pi_\mathrm{sing}$ are cyclic, determined
by the GNS construction from the states $\w_\mathrm{reg}$ and
$\w_\mathrm{sing}$ respectively, then $\pi_r$ is a~cyclic
representation determined by the state\\
 $\w_r(W(J))=\w_\mathrm{reg}\big(\hat{W}(V_\rho)\big)
 \w_\mathrm{sing}\big(w(p(\dV)\oplus r_h(\dV))\big)$.
In particular, we take for $\w_\mathrm{reg}$ the standard vacuum
state,
 $\w_\mathrm{reg}\big(\hat{W}(V_\rho)\big)=
 \exp[-\frac{1}{2}F(\dV_\rho,\dV_\rho)]$, where
\begin{gather}
 \begin{split}
    F(\dV_1,\dV_2)&=-\int_{\w\geq0}
    \ov{\ti{\dV}_1(\w,l)}\cdot\ti{\dV}_2(\w,l)\,\frac{d\w}{\w}\,d^2l\\
    &=\frac{1}{(2\pi)^2}\int\log(s-\tau-i0)\,\dV_1(s,l)\cdot\dV_2(\tau,l)\,
    ds\,d\tau\,d^2l\,,
 \end{split}\\
    \ti{\dV}(\w,l)=\frac{1}{2\pi}\int e^{i\w s}\dV(s,l)\,ds\,.
\end{gather}
Let $B$ be a~positive, trace-class operator in $\Hc_{IR}$ such
that $B^{1/2}\Hc_{IR}$ contains the subspace $C^\infty_{IR}$ of
all smooth fields in $\Hc_{IR}$. Then the formula
\begin{gather}
    \w_\mathrm{sing}\big(w(g\oplus k)\big)=
    \exp\big[-\tfrac{1}{4}s(g\oplus k,g\oplus k)\big]\,,\\
    s(g_1\oplus k_1,g_2\oplus
k_2)=\tfrac{1}{2}(B^{-1/2}g_1,B^{-1/2}g_2)_{\Hc_{IR}}+
2(B^{1/2}k_1,B^{1/2}k_2)_{\Hc_{IR}}\,,
\end{gather}
defines a~quasi-free state. The resulting representation $\pi_r$
satisfies all our selection conditions. If, in addition,
$\ov{B^{-1/2}\C_{IR}^\infty}^{\Hc_{IR}}=\Hc_{IR}$, then $\pi_r$ is
irreducible \cite{her98}.

Representations thus obtained seem to depend on the choice of
$\rho$. However, this dependence is spurious: it was shown in
\cite{her98} that (with fixed $B$) they are all unitarily
equivalent. In fact, one can construct them in a~version which
does not need this auxiliary function. States $\w_r$ with
different $\rho$'s are then realized by different vector states in
the representation space.

The spectrum of energy-momentum covers in each representation from
the given class the whole future lightcone, and is purely
continuous \cite{her98}. Thus there is no vacuum state, but the
energy content can be arbitrarily close to zero. In fact, one
finds that in our formulation the translational invariance is in
contradiction with the regularity of infrared-singular Weyl
operators. However, for infrared-regular test fields the states of
the form discussed above can approximate (weakly) the vacuum.
Indeed, for these test currents $J$ there is $J_\as=0$, so
$J_\rho=J$ and the state $\w_\mathrm{reg}$ is the vacuum. Next, we
have to consider the infrared part $\w_\mathrm{sing}$. As
$p(\dV)=0$, we have
 $\w_\mathrm{sing}\big(\w(0\oplus r_h(\dV))\big)
 =\exp[-\|B^{1/2}r_h(\dV)\|_{IR}^2/2]$. Take now a~family of states
with $\rho_\la(x)=\rho(x+\la t)$, where $t$ is any timelike
vector. It is then easy to show that for $\la\to\pm\infty$ the
corresponding $H_\la$ tends point-wise to $\mp1$, and then
$r_{h_\la}(\dV)$ tends in norm to zero. As the operator $B$ is
bounded, the singular part tends to 1, which ends the proof.

\setcounter{equation}{0}

\section{Scattering}\label{scat}

In standard treatments of scattering the ingoing and outgoing
fields are different representations of one asymptotic field
algebra $\F_\as$ (say, of the free scalar field) unitarily
connected by the scattering operator:
$\pi_\out(A)=S^*\pi_\inc(A)S$, $A\in\F_\as$. The ``in'' and
``out'' fields are (expected to be) obtainable from the actual
field variables of the theory by some limiting process. We have to
explain in what way we expect this picture should be extended to
accommodate two asymptotic algebras $\F_\inc$ and~$\F_\out$.

\subsection{Canonical isomorphism
$\boldsymbol\ga_\io:\F_\mathbf{in}\mapsto\F_\mathbf{out}$}

We still conjecture that the asymptotic fields are outcomes of
some limiting process. The gap between $\F_\inc$ and $\F_\out$
will be bridged by showing that there exists a~canonical
isomorphism $\ga_\io:\F_\inc\mapsto\F_\out$. This automorphism
will be interpreted to describe the ``no interaction'' situation,
in the sense that the formula $\pi_\out(\ga_\io A)=\pi_\inc(A)$,
$A\in\F_\inc$, gives the ``out'' field in terms of the ``in''
field in that case. In general case we expect the relation
\begin{equation}\label{sinout}
    \pi_\out(\ga_\io A)=S^*\pi_\inc(A)S\,,\quad A\in\F_\inc\,,
\end{equation}
with $S$ playing the role of the scattering operator.

To construct the automorphism $\ga_\io$ we first define the group
isomorphism $\io:[\J_\inc]\mapsto[\J_\out]$. To this end we first
define relation $\R\subset[\J_\inc]\times[\J_\out]$ as follows:
$([J_1],[J_2])\in\R$ iff the current $J_1-J_2$ radiates no
electromagnetic field and $(e/4\pi)\int[\Delta V_1(l)-\Delta
V_2(l)]\,d^2l=2k\pi$. Recalling the definition of $[\J_\as]$ as
given after (\ref{ksfsj}) it is easy to see that this is an
unambiguous definition independent of the choice of $J_i$ in the
respective classes, and that the relation is one to one. Moreover,
if the pairs $([J_1],[J_2])$ and $([J'_1],[J'_2])$ satisfy the
relation, then also does the pair $([J_1]+[J'_1],[J_2]+[J'_2])$.
We show in Appendix \ref{inout} that for each $J_1$ there exists
$J_2$ such that $([J_1],[J_2])$ satisfy the relation (and
conversely, for each $J_2$ there is a~respective $J_1$). With this
result we can set $\io[J_1]=[J_2]$ iff $([J_1],[J_2])\in\R$, and
conclude that this defines a~group isomorphism. With the results
of Appendix C it is also easily shown that $\io$ is a~symplectic
mapping, and also that $\{J_1,J_v\}=\{J_2,J_v\}$ for
$[J_2]=\io[J_1]$.

We now set
\begin{equation}\label{isga}
    \ga_\io\big(\psi_\inc(\chi)\big)=\psi_\out(\chi)\,,\quad
    \ga_\io\big(W_\inc(J_1)\big)=W_\out(J_2)\,,\quad [J_2]=\io[J_1]\,.
\end{equation}
If $\pi_\inc$ is the representation of $\B_\inc$ in the assumed
class (see (\ref{rep}) and the accompanying discussion) then
$\pi_\out$ defined by
$\pi_\out(\psi_\out(\chi))=\pi_\inc(\psi_\inc(\chi))$,
\mbox{$\pi_\out(W_\out(J_2))=\pi_\inc(W_\inc(J_1))$},
$[J_2]=\io[J_1]$, defines a~representation of $\B_\out$ in the
same class. Therefore $\ga_\io$ extends to a~topological
isomorphism $\ga_\io:\F_\inc\mapsto\F_\out$.

\subsection{Radiation by external current}

How to construct a~scattering theory in the given language in the
general case of full quantum theory is an open question. Here we
consider only electromagnetic field scattered by classical
external current.

Let $J_\ext$ be a~current satisfying the conditions of spaces
$\J_\as$, except that its asymptote has both an incoming and
outgoing parts (has support in $\ov{V_+\cup V_-}$); this is a
classical conserved current typical of charged matter. We denote
\mbox{$V_\ext(s,l)=\int J_\ext(x)\delta(s-x\cdot l)dx$}. This
current produces the Lorenz radiation potential $A_\ext$ in
accordance with (\ref{rad}). We need test currents producing the
same radiation as $J_\ext$ but belonging to $\J_\inc$ or
$\J_\out$. Let $J_{\ext,\inc}$ and $J_{\ext,\out}$ denote the
incoming and outgoing asymptote of $J_\ext$ respectively. We
denote further
\begin{equation}\label{extas}
 \begin{split}
    J_{\ext,\out}^+(x)&=J_{\ext,\out}(x)+J_{\ext,\out}(-x)\,,
    \quad
    J_{\ext,1}=J_\ext-\rho*J_{\ext,\out}^+\,,\\
    J_{\ext,\inc}^+(x)&=J_{\ext,\inc}(x)+J_{\ext,\inc}(-x)\,,
    \quad\quad
    J_{\ext,2}=J_\ext-\rho*J_{\ext,\inc}^+\,.
 \end{split}
\end{equation}
Then proceeding as in Appendix \ref{inout} one shows that
$J_{\ext,1}\in\J_\inc$, $J_{\ext,2}\in\J_\out$ and they both
produce radiation potential $A_\ext$; the classes $[J_{\ext,1}]$
and $[J_{\ext,2}]$ are independent of the choice of $\rho$ and
$[J_{\ext,2}]=\iota[J_{\ext,1}]$. Denoting\\
 $V_{\ext,i}(s,l)=\int J_{\ext, i}\delta(s-x\cdot l)dx$ ($i=1,2$)
we have\\ $V_{\ext,1}(s,l)=V_\ext(s,l)-V_\ext(+\infty,l)$,
$V_{\ext,2}(s,l)=V_\ext(s,l)-V_\ext(-\infty,l)$.

The process of scattering of electromagnetic field by the external
classical current should have the effect of adding the radiation
field of the current. Therefore, remembering the interpretation of
$W(J)$ we expect
\begin{equation}\label{scatext}
    \pi_\out(W_\out(J_2))=\pi_\inc(W_\inc(J_1))
    e^{-i\{J_2,J_{\ext,2}\}}=
    \pi_\inc(W_\inc(J_1))
    e^{-i\{J_1,J_{\ext,1}\}}\,,
\end{equation}
where $[J_2]=\iota[J_1]$. Using the commutation relations it is
easy to find that the operator
\begin{equation}\label{S}
    S=\pi_\inc(W_\inc(J_{\ext,1}))=\pi_\out(W_\out(J_{\ext,2}))
\end{equation}
satisfies the condition  (\ref{sinout}) and turns to identity when
there is no radiation. There is no ``infrared catastrophe''
difficulty in this formulation.

Formula (\ref{S}) cannot be directly applied when $J_\ext$ is a
current of a~point charge,
\begin{equation}\label{point}
    J_\mathrm{point}(x)=e\int \dot{z}(\tau)
    \delta^4(x-z(\tau))\,d\tau\,.
\end{equation}
However, if $\pi_\inc$ (and $\pi_\out$) is of the form discussed
in Section \ref{alg-ex} it can be extended to this case. One
easily finds that
\begin{equation}\label{vpoint}
    V_\mathrm{point}(s,l)=
    e\frac{\dot{z}(\tau(s,l))}{\dot{z}(\tau(s,l))\cdot l}\,,\quad
    V_{\mathrm{point}\rho}(s,l)=
    \int \dot{H}(s-u,l)V_\mathrm{point}(u,l)\,du\,,
\end{equation}
where $\tau(s,l)$ is the solution of $z(\tau)\cdot l=s$. For a
smooth trajectory $z(\tau)$ with sufficiently fast achieved
asymptotic velocities the element
$\hat{W}(V_{\mathrm{point}\rho})$ is well defined and the formula
(\ref{S}) may be extended with the use of the rhs of (\ref{pir}).

\setcounter{equation}{0}

\section{Summary}\label{sum}

This paper introduces the notion of spatially local observables
and fields in electrodynamics; we have developed the test
functions machinery needed for that. We have shown that the
enlarged algebra of fundamental asymptotic fields naturally
includes the infrared degrees of freedom, which care for the
implementation of Gauss' law at the algebraic level. The algebra
proved to be a~reformulation of the algebra postulated before by
the author, but now we gained control over the spacetime
localization of fields.

The perturbational construction of the standard electrodynamics is
based on the uncoupled free fields algebra. We expect that the
scattering theory based in similar way on the algebra discussed in
this article could throw new light on the infrared and charge
problems of the standard theory. Here, as a~first step towards
this task, we have shown how to apply our formalism to the simple
case of the radiation produced by external classical current. The
description is free from ``infrared catastrophe''.

\setcounter{equation}{0}

\section*{Appendices}
\appendix

\section{Functions $F_\pm(p)$}\label{Fplmi}

Let $V_a(l)$ be a~smooth vector function on the future lightcone,
homogeneous of degree $-1$ and such that $l\cdot V(l)=0$. If $t$
is any unit timelike future-pointing vector and $V(l)$ is
continued to a~neighbourhood of the cone with the preservation of
its properties then $L_{ab}[t^aV^b(l)/t\cdot l]=\p\cdot V(l)$.
Therefore the rhs is independent of the extension in the assumed
class and
\begin{equation}\label{pV}
    \int\p\cdot V(l)\,d^2l=0\,.
\end{equation}

Let now $F(w)$ be defined for $w=p+iq$, $p,q\in M$, $q^2>0$, by
\begin{equation}\label{F}
    F(w)=\int\frac{w\cdot V(l)}{w\cdot l}\,d^2l\,.
\end{equation}
This is a~homogeneous of degree $0$, analytical function on its
domain. Choosing any unit timelike future-pointing vector $t$ and
using property (\ref{pV}) we rewrite $F$ as
\begin{equation}\label{Flog}
    F(w)=-\int\p\cdot V(l)\log\frac{w\cdot l}{t\cdot l}\,d^2l+
    \int\frac{t\cdot V(l)}{t\cdot l}\,d^2l\,.
\end{equation}
This is used to show that $F(w)$ is bounded on its domain. Due to
homogeneity it is sufficient to consider two cases:
$|q|\leq|p|=1/2$, and $|p|\leq|q|=1/2$. In the first case the
first integral is bounded by
$\con\int\big[|\log(p^0-\vec{p}\cdot\vec{l})^2|+2\pi\big]
d\Omega(\vec{l})\leq\con$; the second case is treated similarly.

It is easy to see that there exist limit functions on
$M\setminus\{0\}$
\begin{multline}\label{Fext}
    F_\pm(p)=\lim_{\la\searrow0}F(p\pm i\la t)\\
    =-\int\p\cdot V(l)\Big[\log\frac{|p\cdot l|}{t\cdot l}
    \mp i\frac{\pi}{2}\sgn(p\cdot l)\Big]\,d^2l+
    \int\frac{t\cdot V(l)}{t\cdot l}\,d^2l=F_\mp(-p)\,,
\end{multline}
which are independent of the choice of $t$ (specified as above),
homogeneous and equal to $F(p)$ for $p\in V_\pm$. We show below
that these limits are achieved uniformly on each set separated
from zero, $F_\pm$ are continuous outside $p=0$ and $\C^\infty$
outside $p^2=0$, and for each multiindex $\al$ functions
$(p^2)^{|\al|}D^\al F_\pm(p)$ have continuous extensions to $M$.

\noindent
 {\bf Proof}

If $g(l)$ is any smooth function on the cone, homogeneous of
degree $-2$ and such that $\int g(l)\,d^2l=0$, then for $q^2>0$ we
have
\begin{multline}\label{pwg}
    \frac{\p}{\p w^a}
    \int g(l)\log\frac{w\cdot l}{t\cdot l}\,d^2l\\
    =\frac{w^b}{w^2}
    \bigg[\int L_{ba}g(l)\log\frac{w\cdot l}{t\cdot l}\,d^2l
    +\int g(l)\frac{l_at_b-l_bt_a}{t\cdot l}d^2l\bigg]\,.
\end{multline}
This is shown by transferring $L_{ba}$ on the rhs by parts and
observing that
\begin{equation*}
 [L_{ba}+W_{ba}]\log(w\cdot l/t\cdot l)=
 (l_at_b-l_bt_a)/t\cdot l\,,
\end{equation*}
where $W_{ba}=w_b\,\p/\p w^a-w_a\,\p/\p w^b$. Applying (\ref{pwg})
inductively to (\ref{Flog}) one finds that for $q^2>0$
\[
 (w^2)^{|\al|}D^\al F(w)=\sum_iQ_i(w)
 \int h_i(l)\log\frac{w\cdot l}{t\cdot l}\,d^2l
 +Q(w)\,,
\]
where $Q_i$ and $Q$ are polynomials, homogeneous of degree
$|\al|$, and $h_i$ are smooth functions, homogeneous of degree
$-2$ and such that $\int h_i(l)d^2l=0$. To end the proof it is now
sufficient to show that uniform limits (\ref{Fext}) exist on each
set separated from zero for each function of the form $G(w)=\int
h(l)\log(w\cdot l/t\cdot l)\,d^2l$, with $\int h(l)d^2l=0$. We
have ($\la>0$)
\[
 \log\frac{(p\pm i\la t)\cdot l}{t\cdot l}
 =\tfrac{1}{2}\log\bigg[\Big(\frac{p\cdot l}{t\cdot l}\Big)^2+\la^2\bigg]
 \mp i\arctan\Big(\frac{p\cdot l}{\la\, t\cdot l}\Big)\pm i\frac{\pi}{2}\,,
\]
but the last term falls out of the integral. Denote
\[
 G_\pm(p)=\int h(l)\Big[\log\frac{|p\cdot l|}{t\cdot l}
    \mp i\frac{\pi}{2}\sgn(p\cdot l)\Big]\,d^2l\,.
\]
If we denote $k=p/\la$ then
\begin{multline*}
 |G(p\pm i\la t)-G_\pm(p)|=|G(k\pm it)-G_\pm(k)|\\\leq\con
 \int\bigg\{\log\bigg[1+\Big(\frac{t\cdot l}{k\cdot l}\Big)^2\bigg]
 +\bigg[\pi
 -2\arctan\frac{|k\cdot l|}{t\cdot l}\bigg]\bigg\}\,d^2l\\
 =\frac{\con}{|\vec{k}|}
 \int_{|k^0|-|\vec{k}|}^{|k^0|+|\vec{k}|}
 \Big\{\log\Big[1+\frac{1}{u^2}\Big]+\pi-2\arctan|u|\Big\}du\\
 \leq\con\bigg\{\log\bigg[1+\frac{1}{|k^0|+|\vec{k}|}\bigg]
 +\frac{\log[1+|k^0|+|\vec{k}|]}{|k^0|+|\vec{k}|}\bigg\}\,.
\end{multline*}
where the spherical coordinates have been used. To get this result
we first estimate the integrand by $\con\log(1+1/|u|)$ and then
consider the cases \mbox{(a) $|k^0|\geq2|\vec{k}|$} and (b)
$|k^0|\leq2|\vec{k}|$ separately. In case (a) the integral is
bounded by \mbox{$\con\log[1+1/(|k^0|-|\vec{k}|)]$}, which is
sufficient as here $|k^0|-|\vec{k}|\geq(|k^0|+|\vec{k}|)/3$. In
case (b) we extend the integration limits to
$\pm(|k^0|+|\vec{k}|)$, calculate the integral explicitly and
observe that here $|\vec{k}|\geq(|k^0|+|\vec{k}|)/3$, which yields
the final result.

For $|p^0|+|\vec{p}|\geq r\geq\la$ we now have
\[
    |G(p\pm i\la t)-G_\pm(p)|\leq\con
    \frac{1+\log[1+r/\la]}{r/\la}\to0\quad
    \text{for}\quad\la\to0\,,
\]
which ends the proof.

\setcounter{equation}{0}

\section{Transformation $\boldsymbol E^F$}\label{phase}

Let $F$ and $F_\pm$ be defined as in Appendix \ref{Fplmi} and let
$\chi\in\Sb_k(\ov{\C_\pm})$ for a~given $k\geq10$. Define linear
mappings $\chi\mapsto E^F_\pm\chi$ by
\begin{equation}\label{WF}
    \widehat{E^F_\pm\chi}(p)=\exp[-iF_\pm(p)]\hat{\chi}(p)\,.
\end{equation}
We show here that there is $E^F_\pm\chi\in\Sb_{k-5}(\ov{\C_\pm})$.
Furthermore, if $\chi\in\Sb_k(\ov{\C})$ and $\C=\C_+\cup\C_-$ then
it is possible to separate $\chi=\chi_{1+}+\chi_{1-}$,
$\chi_{1\pm}\in\Sb_k(\ov{\C_\pm})$, and define
\mbox{$E_1^F\chi=E_+^F\chi_{1+}+E_-^F\chi_{1-}\in\Sb_{k-5}(\ov{\C})$};
subscript $1$ indicates the choice of (in general non-unique)
separation. If another separation is indexed by $2$ then
\linebreak $\wh{E^F_1\chi}(p)=\wh{E^F_2\chi}(p)$ for $p^2>0$.

\noindent
 {\bf Proof}

If $\chi\in\Sb_k$ ($k\geq5$) then one shows by induction with
respect to $|\alpha|$ that $\hat{\chi}(p)$ is $\C^\infty$ outside
$p^2=0$ and for $|\al|\leq k-5+n$ the functions
$(p^2)^nD^\al\hat{\chi}(p)$ have continuous extensions to $M$,
vanishing faster than polynomially at infinity. Using this fact
and the properties of $F_\pm$ shown in Appendix \ref{Fplmi} it is
easy to show that the same remains valid when $\hat{\chi}(p)$ is
replaced by $\hat{\chi'}(p)=\exp[-iF_\pm(p)]\hat{\chi}(p)$. It
follows that $\sup_x|x^\al\Box^nD^\be\chi'(x)|<\infty$ whenever
$|\al|\leq(k-5)+n$. This is sufficient to conclude that
$E^F_\pm\chi\in\Sb_{k-5}$.

Let $\chi\in\Sb_k(\ov{\C_\pm})$, $k\geq10$, so
$E^F_\pm\chi\in\Sb_{k-5}$. We can choose the reference point at
the vertex of the cone, and then the support of $\chi$ is in
$\ov{V_\pm}$. Then the Fourier transform $\hat{\chi}(w)$ exists as
an analytical function of the complex variable for\linebreak $w\in
M+ iV_\pm$, satisfies the bound
$|\hat{\chi}(p+iq)|<\con(1+|p|)^{-n}$ for each $n$ and gives
$\dsp\lim_{\la\searrow0}\hat{\chi}(p+i\la q)=\hat{\chi}(p)$. Using
the properties of $F(w)$ one finds that also
$\exp[-iF(w)]\hat{\chi}(w)$ is analytical in the same domain,
satisfies similar bounds and gives in the limit
$\dsp\lim_{\la\searrow0}\exp[-iF(p+i\la q)]\hat{\chi}(p+i\la
q)=\exp[-iF_\pm(p)]\chi(p)$. By the theorem connecting cone-like
support properties of a~distribution with the analyticity
properties of its transform (see \cite{ree}, Thm.IX.16) this is
sufficient to conclude that
$E^F_\pm\chi\in\Sb_{k-5}(\ov{\C_\pm})$.

Let now $\chi\in\Sb_k(\ov{\C})$, $\C=\C_-\cup\C_+$. As
$\C_-\cap\C_+\neq\emptyset$ one can choose the reference system
such that $\C_\pm=\mp Re_0+V_\pm$, $R>0$ (if the origin is
identified with the zero vector). Let $f$ be a~real smooth
function of a~real variable, such that $f(s)=-1$ for $s<-1/3$,
$f(s)=1$ for $s>1/3$. We define functions $\rho_\pm$ on $\M$ as
follows:
\begin{equation}
 \rho_\pm(x)=
 \begin{cases}
   \dsp\frac{1}{2}\bigg[1\pm
   f\Big(\frac{Rx^0}{R^2-|\vec{x}|^2}\Big)\bigg]&\text{for}\quad
   x\in\C_-\cap\C_+\\
   \dsp1&\text{on the rest of}\ \C_\pm\\
   \dsp0&\text{outside}\ \C_\pm\,.
 \end{cases}
\end{equation}
It is easy to show that $\rho_\pm\in\C^\infty(\C)$ and the
functions $\chi_\pm=\rho_\pm\chi$ are in $\Sc_k^\Box(\ov{\C_\pm})$
respectively (use the fact that $\chi$ and its derivatives vanish
at the boundary of $\C$ faster than any power of the Euclidean
distance from that boundary). As the sum $\rho_++\rho_-$ is the
characteristic function of the set $\C$ we have
$\chi=\chi_++\chi_-$ and this separation satisfies conditions
stated at the beginning. If $\chi=\chi_{i+}+\chi_{i-}$, $i=1,2$,
are any two separations satisfying these conditions, then
$\wh{E_2^F\chi}(p)-\wh{E_1^F\chi}(p)=
\big(\exp[-iF_+(p)]-\exp[-iF_-(p)]\big)
\big(\hat{\chi}_{2+}(p)-\hat{\chi}_{1+}(p)\big)$, which vanishes
for $p^2>0$.

\setcounter{equation}{0}

\section{Non-radiating currents}\label{inout}

Let $J_1\in\J_\inc(\ov{\C})$, $\C=\C_-\cup\C_+$, with the
asymptote $J_{1\as}$. Choose a~real smooth function $\rho$ on $\M$
with support in $\C_-\cap\C_+$, such that $\int \rho(y)dy=1$, and
define
\begin{equation}\label{itoo}
 \begin{split}
    J_2=J_1-\rho*J_{1\as}^+\,,\quad
    (\rho*J_{1\as}^+)(x)&=\int \rho(x-y)J_{1\as}^+(y)\,dy\,,\\
    J_{1\as}^+(y)&=J_{1\as}(y)+J_{1\as}(-y)\,.
 \end{split}
\end{equation}
Then we have $J_2\in\J_\out(\ov{\C})$, $J_2-J_1$ produces no
radiation potential and\linebreak $J_{2\as}(x)=-J_{1\as}(-x)$. In
terms of $V$'s we have $V_2(s,l)=V_1(s,l)-V_1(-\infty,l)$ and
$\Delta V_2(l)=V_2(+\infty,l)=-V_1(-\infty,l)=\Delta V_1(l)$.

\noindent
 {\bf Proof}

It is easy to show that the asymptote $J_{1\as}(x)$ is
divergence-free outside $x=0$ (as a~limit of a~conserved current).
But it is also a~distribution and by smearing it with a~test
function one also shows that $J_{1\as}$ is a~conserved
distributional current. Then it is immediate that
$\rho*J_{1\as}^+$ is a~smooth, conserved current. The support
properties of $\rho$ and $J_1$ imply that the support of
$\rho*J_{1\as}^+$, and consequently also the support of $J_2$, are
contained in $\ov{\C}$.

Let $\C_-\cap\C_+$ be contained in $|x|\leq R$. Then for $|x|\geq
2R$ we have
\[
 |D^\be H_3(\rho*J_{1\as}^+)(x)|=\Big|\int \rho(z)\,z\cdot\p D^\be
 J_{1\as}^+(x-z)dz\Big|\leq\con\,|x|^{-|\be|-4}\,.
\]
This guarantees that $\rho*J_{1\as}^+,J_2\in\Sc^3_{3+}(\ov{\C})$
together with $J_1$. The asymptote of $\rho*J_{1\as}^+$ is easily
found: $\lim_{\la\to\infty}\la^3(\rho*J_{1\as}^+)(\la
x)=J_{1\as}^+(x)$ (use the assumption on the integral of $\rho$).
Thus $J_{2\as}(x)=-J_{1\as}(-x)$ and $J_2\in\J_\out(\ov{\C})$.

Finally, we calculate
\begin{multline*}\label{vf}
    V_1(s,l)-V_2(s,l)=\int\delta(s-x\cdot l)(\rho*J_{1\as}^+)(x)dx\\
    =\lim_{\la\searrow0}\int_{|s-z\cdot l|\geq\la}\rho(z)\bigg\{\int
    \delta(s-z\cdot l-y\cdot l)J_{1\as}^+(y)dy\bigg\}dz\,.
\end{multline*}
But now using the fact that $|\xi|^3J_{1\as}^+(\xi
y)=J_{1\as}^+(y)$ for all $\xi\neq0$ one can scale $s-z\cdot l$ to
$-1$, and then one finds
\[
 V_1(s,l)-V_2(s,l)=\int\delta(y\cdot l+1)J_{1\as}^+(y)dy=
 \int\delta(y\cdot l+1)J_{1\as}(y)dy=V_1(-\infty,l)\,,
\]
which ends the proof.

\setcounter{equation}{0}

\end{document}